\documentclass[twocolumn,twocolappendix]{aastex631}

\usepackage{placeins}
\usepackage{appendix}
\usepackage{graphicx}
\graphicspath{{Figures_apj_letter/}}
\begin{document}

\title{A Detailed Polarimetric Imaging Study of Type II Solar Radio Burst with the MWA: Detection of Linear Polarization}

\author[0000-0002-0786-7307]{Puja Majee}
\affiliation{National Centre for Radio Astrophysics, Tata Institute of Fundamental Research, Pune University Campus, Pune, India, 411007}

\author[0000-0001-8801-9635]{Devojyoti Kansabanik}
\affiliation{Instituto de Astrofísica de Andalucía (IAA-CSIC), Glorieta de la Astronomía s/n, E-18008, Granada, Spain}
\affiliation{Johns Hopkins University Applied Physics Laboratory, 11001 Johns Hopkins Rd, Laurel, MD, USA}
\affiliation{University Corporation for Atmospheric Research, 3090 Center Green Dr., Boulder, CO, USA}

\author[0000-0002-4768-9058]{Divya Oberoi}
\affiliation{National Centre for Radio Astrophysics, Tata Institute of Fundamental Research, Pune University Campus, Pune, India, 411007}

\begin{abstract}
Type II radio bursts, produced by magnetohydrodynamic shocks propagating through the solar corona and interplanetary medium, serve as powerful diagnostics of shock properties and ambient coronal magnetic fields. While circular polarization in type II bursts has been extensively studied, most investigations rely on dynamic spectra (DS), providing only spatially integrated measurements. DS obscures spatial variations of the polarization sign and magnitude, potentially causing cancellation effects and misrepresenting true polarization levels.
We present a full-polarimetric imaging study of a metric type II burst observed with the Murchison Widefield Array at 78–133 MHz. Remarkably, we report the first detection of linear polarization in a metric type II burst, reaching a polarization fraction of $\sim12\%$ with significant spectro-temporal variability across observing bands. 
The spectro-temporal, spatial, and morphological properties found in this study establish linear polarization as a new diagnostic of large-scale coronal and interplanetary magnetic field structure through full-Stokes imaging of metric solar radio bursts -- challenging conventional paradigms predicting negligible linear polarization from coherent plasma emission processes. 
\end{abstract}


\section{Introduction} \label{sec:intro}
Type II solar radio bursts are radio signatures of magneto-hydrodynamic (MHD) shocks in the corona and the interplanetary medium. As the shock propagates through the corona, it accelerates the electrons in the ambient coronal medium to supra-thermal energies, which excite the Langmuir waves that convert into electromagnetic emissions in the radio band through non-linear processes. This emission mechanism, known as plasma emission, emits at the local plasma frequency and its harmonics \citep{Nelson_Melrose1985_typeII_burst}.
In dynamic spectra, plasma emissions from type II radio bursts typically appear as two slowly drifting emission bands at the fundamental (F) and harmonic (H) of the local plasma frequency at the site of emission. Both F and H emission bands are often observed to be split into two or sometimes more lanes, referred to as band-split and multilane type IIs, respectively \citep{Nelson_Melrose1985_typeII_burst}.

Type II bursts serve as one of the few remote-sensing probes for the shock properties --  shock compression ratio, shock speed, Alfv$\acute{\text{e}}$nic mach number -- through the middle corona to the inner heliosphere, the heights where formation and early evolution of the shocks take place \citep[][etc]{Smerd1975, Vrsnak2001, Vrsnak2002, Nindos2011, Maguire2020}. Beyond kinematics, their polarimetric properties offer a valuable diagnostic for estimating the strength of the shock-entrained magnetic field \citep{Hariharan2014, Kumari2017typeII_band, Ramesh_and_Kathiravan2022, Ramesh_2023}. 

While theories of plasma emission predict the fundamental emission to be strongly circularly polarized up to $100\%$ and harmonic emission to be weakly circularly polarized \citep{Melrose1977, MelroseSy1972, Melrose1980, Zlotnik1981}, type II bursts are reported to exhibit a wide range of circularly polarized emission, ranging from weakly or negligibly polarized \citep[][etc]{Komesaroff1958, roberts1959solar, Hariharan2014,Ramesh_and_Kathiravan2022, Ramesh_2023} to strongly polarized \citep[][etc]{Thejappa2003, Du2014, Alissandrakis2021}. The observed low degree of polarization is attributed to propagation effects, including reflection off sharp boundary layers in corona \citep{Melrose_2006}, mode coupling \citep{cohen1960,melrose_1974}, and scattering due to coronal inhomogeneities along the ray path \citep{Melrose_1989}.

The majority of these studies rely on Sun-as-a-star observations, providing spatially averaged information. The polarized emission can be positive and/or negative depending on the sense of polarization, and also can vary across the spatial extent of the type II source. Hence, spatially integrated observations can lead to misleading measurements of polarized emission. This underscores the importance of imaging in polarimetric studies.

Linear polarization from a type II solar radio burst has not been reported yet. This is largely consistent with theoretical expectations, as the plasma emission mechanisms predict circularly polarized radiation. Besides, any linear polarization generated at the source is expected to be destroyed by the large Faraday rotation (FR) encountered during coronal propagation \citep{Grognard1973}.
As a consequence, for the past many decades, any signature of linearly polarized emission from solar radio emission has been attributed to instrumental leakage and often has been removed as part of the polarization calibration \citep{Grognard1973, Kumari2017typeII_band, Morosan2022, Ramesh_2023}. 
This conventional wisdom has recently been challenged by the imaging detection of linearly polarized emission from type I and III radio bursts reported by \citet{Dey_2025_linear}.
There is only other report of imaging detection of linearly polarized emission from the Sun - by \citet{Alissandrakis1994_linear_pol} at 5 GHz.
We note that in the past few years, there have been a few reports of elliptically polarized stellar burst emissions, including a stellar type II burst \citep{Callingham2021, Bastian2022, Callingham2025_stellar_typeII}.

This letter presents the first detailed polarimetric imaging study of a type II burst. Not only is this the first such study from the Murchison Widefield Array \citep[MWA;][]{Tingay2013_mwa, Wayth2018_mwa}, it is also the first to report linear polarization from a type II burst.
The primary focus of this letter is the characterization of linearly polarized emission from the type II; a companion paper will present a comprehensive study of the circularly polarized emission from the same event. This letter is structured as follows: Section \ref{event-details} describes the type II event observed by the MWA; Section \ref{data-analysis} outlines the calibration and imaging methodologies along with the image analysis. The results from this study and their discussions are presented in Sections \ref{sec:results} and \ref{sec:discussion}, respectively, followed by the conclusions in Section \ref{sec:conclusion}.

\begin{figure*}[!htbp]  
    \centering
    \includegraphics[width=0.48\linewidth]{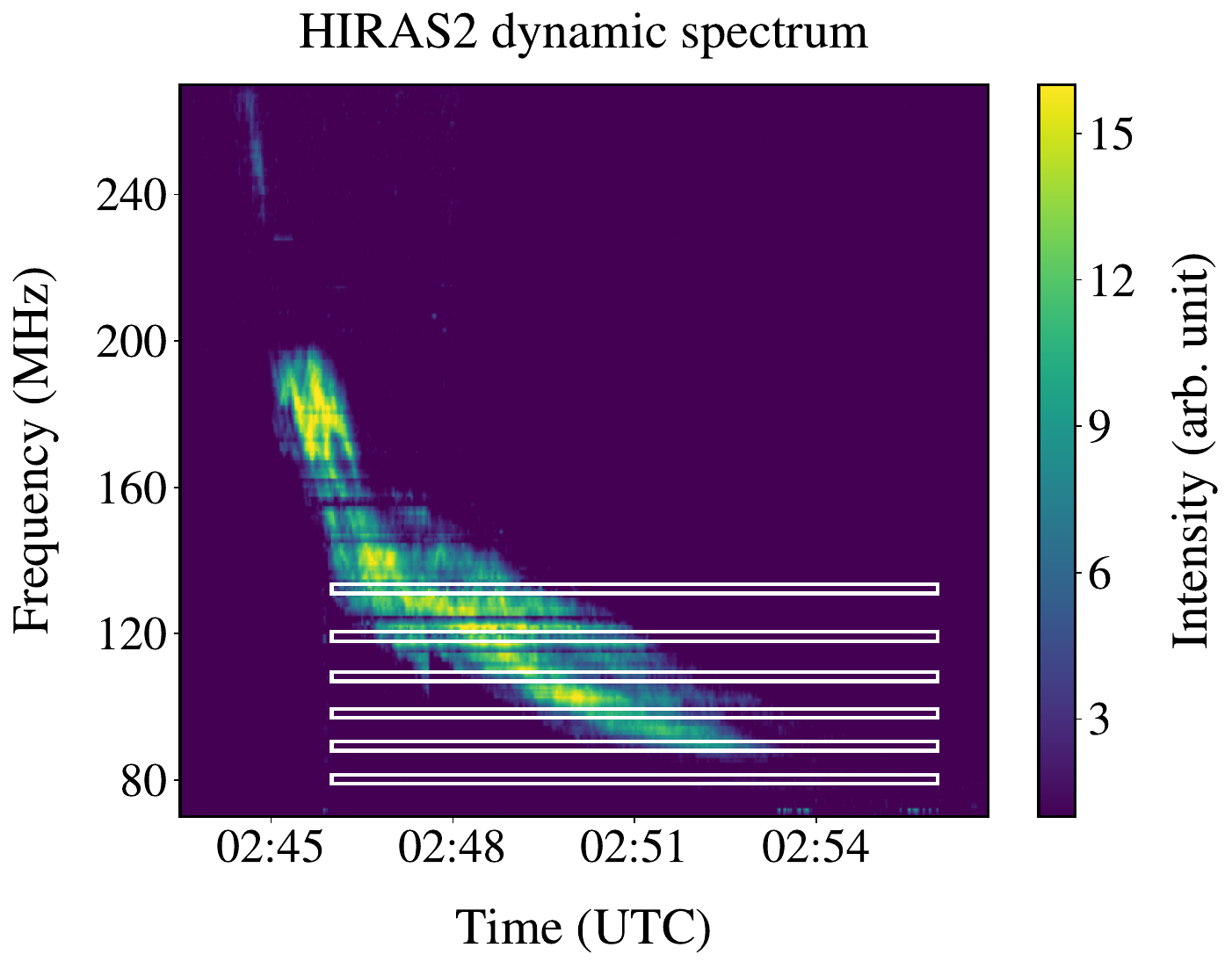}
    \includegraphics[width=0.50\linewidth]{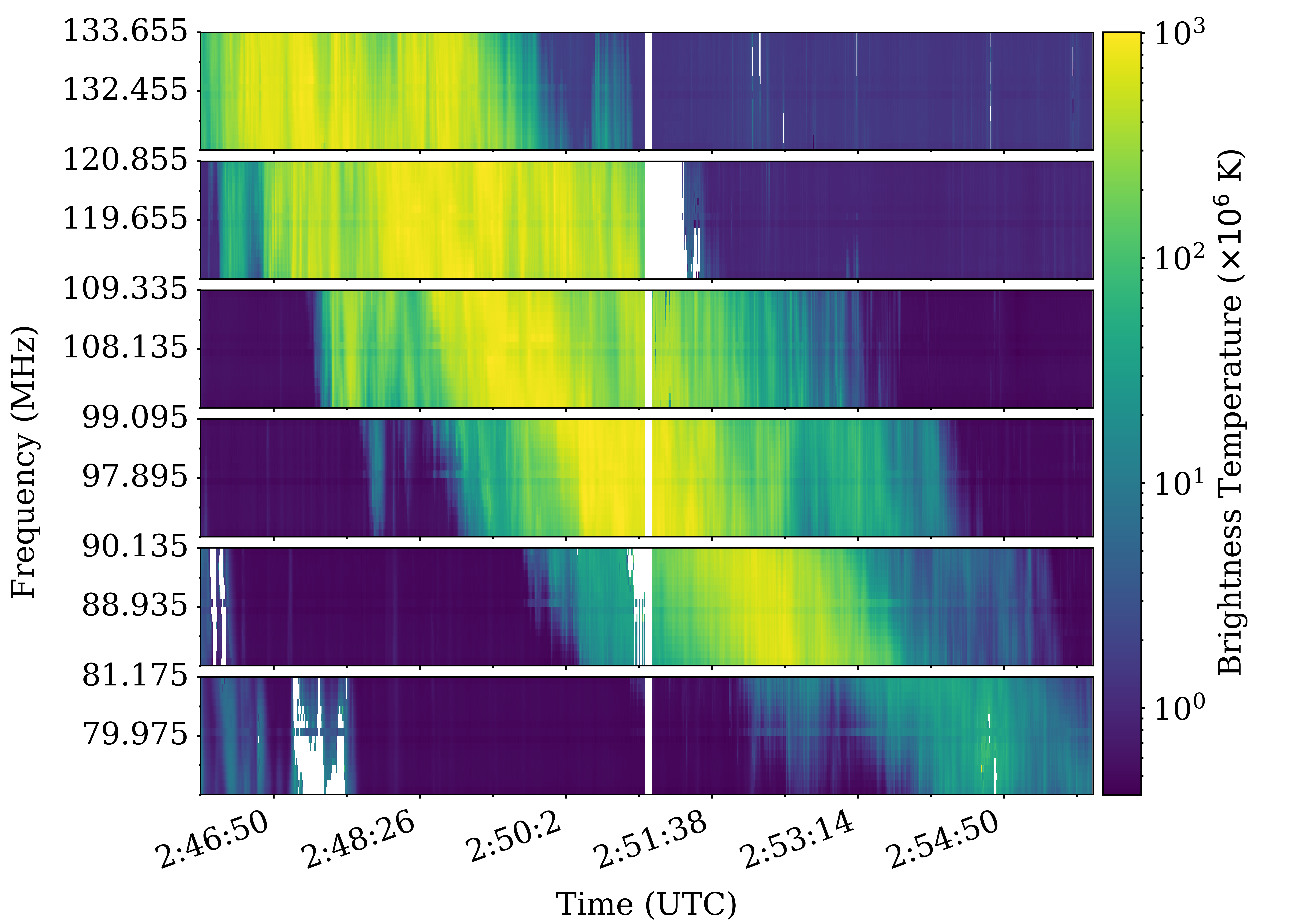} 
    \caption{Left panel: Dynamic spectrum of the type II harmonic emission captured by the HIRAS2 spectrograph \citep{Kondo1995}, with the color representing the intensity of the type II burst in arbitrary units. The six white rectangles mark the spectral bands, each spanning 2.56 MHz, corresponding to the spectral and temporal coverages of the MWA observation. Right panel: Dynamic spectrum of the type II emission observed by the MWA across the six spectral bands. The color scale represents the peak brightness temperature at the location of the type II burst} determined from spectroscopic snapshot images. The white vertical strip at the centre of each spectral band represents the time gap between the two successive observations of the event, while the other white regions correspond to bad images.
    \label{letter_fig:dynamic_spectra_and_brightness_temperature}
\end{figure*}

\section{Type-II Event Details}\label{event-details}
This type II burst was observed by the MWA on 28th September, 2014, between 2:46:00 and 2:55:52 UTC. At that time, the MWA was operating in its Phase 1 configuration \citep{Tingay2013_mwa}, consisting of 128 array elements, referred to as tiles, distributed across approximately 3 km in a centrally condensed array layout. The data has a temporal resolution of 0.5 s and a spectral resolution of 40 kHz. The array provides an instantaneous bandwidth of 30.72 MHz, divided into 24 coarse spectral channels of 1.28 MHz each — placed either in a discrete or a contiguous fashion — spanning an operational frequency range of 80 to 300 MHz. 
During the observation of this type II event, only 12 out of 24 coarse channels yielded usable data due to instrumental issues, spanning frequency range 78 to 133 MHz, which are shown in the left panel of Figure \ref{letter_fig:dynamic_spectra_and_brightness_temperature} as six rectangular boxes each covering 2 coarse channels. The right panel shows MWA's observation of the type burst across the six spectral bands, expressed in units of brightness temperature. The event was captured in two consecutive 4-minute segments.
More detailed information about this event can be found in \citet{Bhunia_2023}, where the authors carried out a detailed total intensity (Stokes I) spectroscopic snapshot imaging study of this type II event. 

\begin{figure*}[!htbp]  
    \centering
    \includegraphics[width=0.49\linewidth]{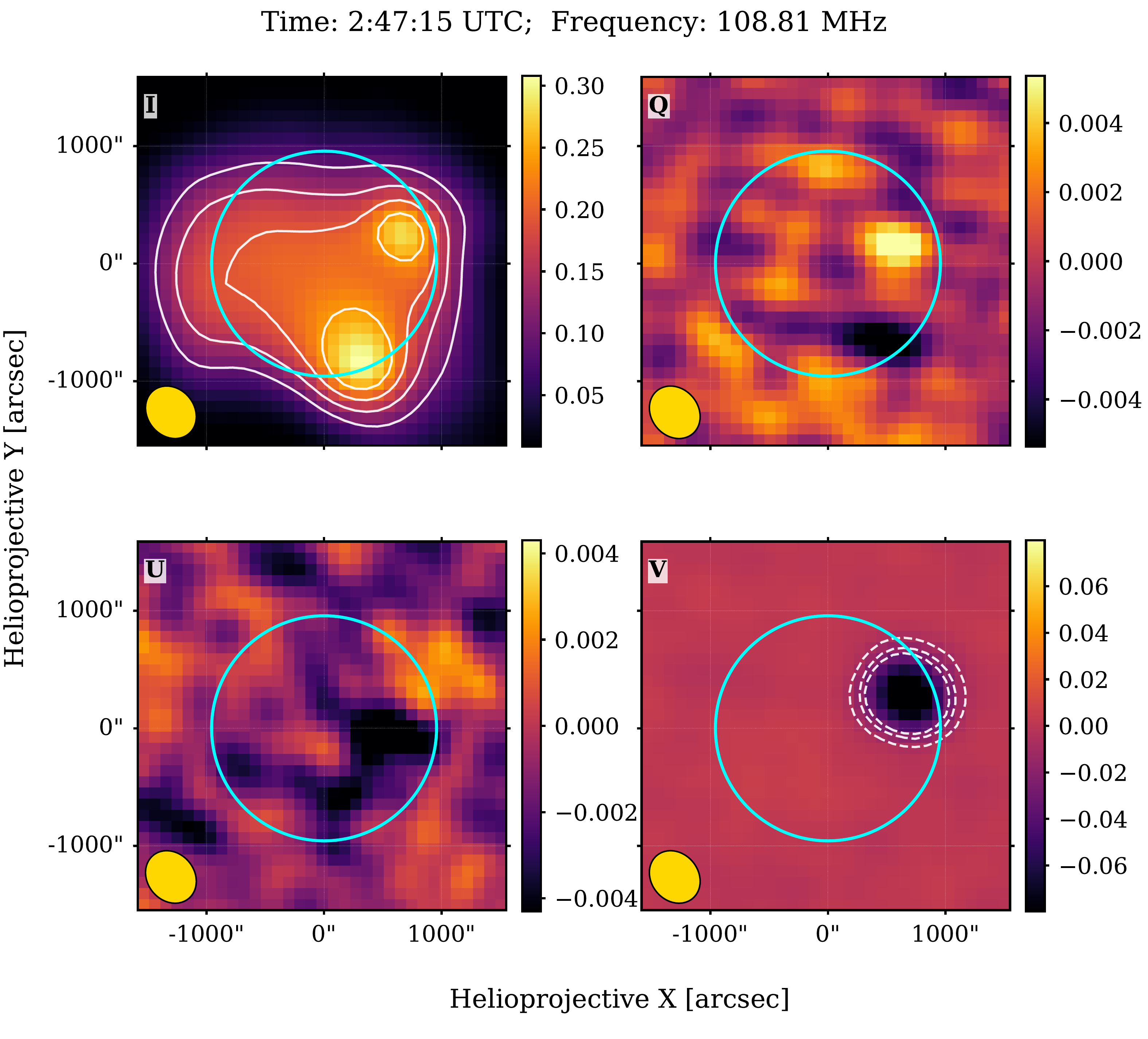}
    \includegraphics[width=0.49\linewidth]{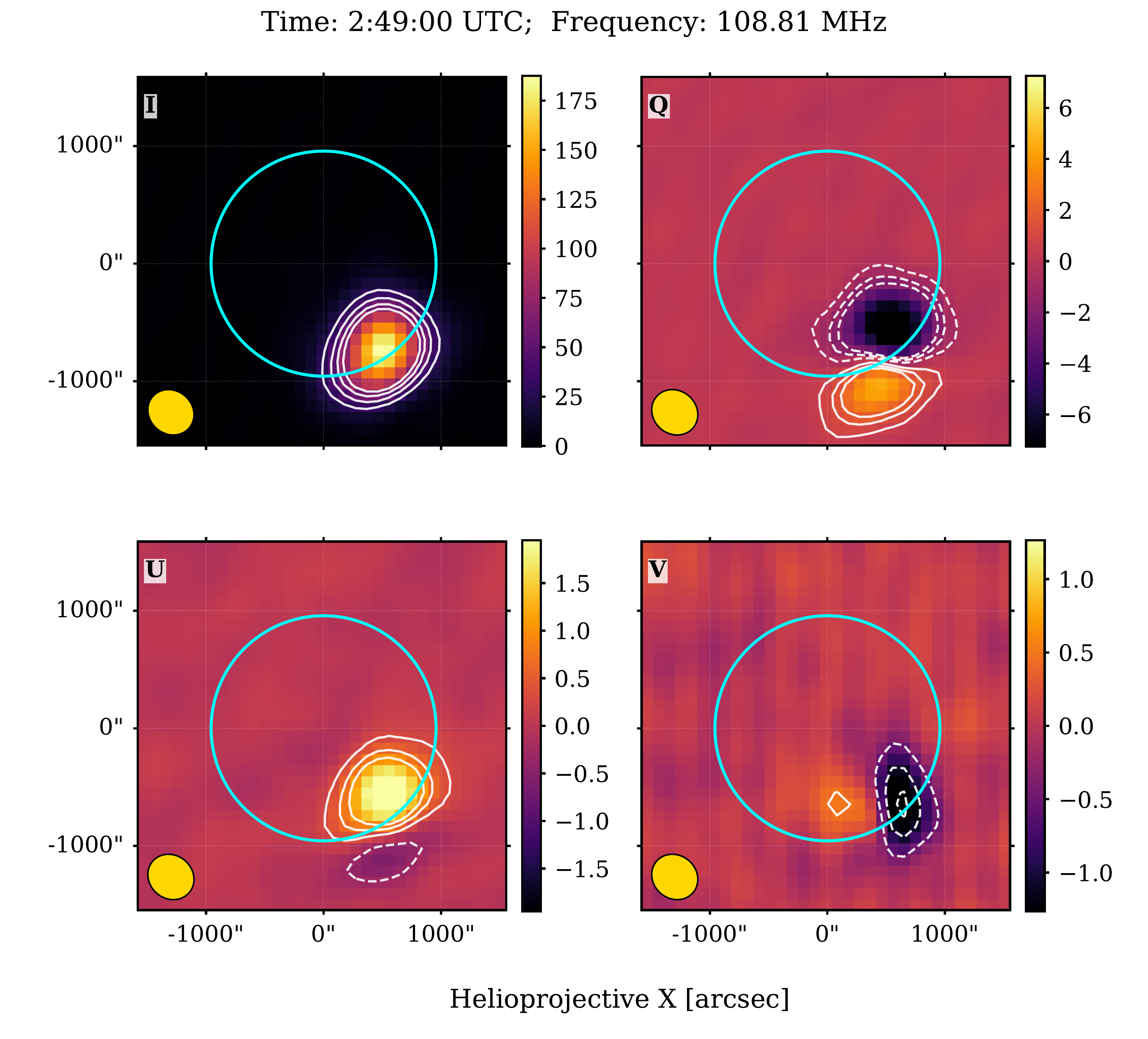}
    \caption{Two sets of full Stokes maps of the Sun in the helioprojective coordinates at 108.81 MHz are shown -- Stokes I (upper left), Q (upper right), U (lower left), and V (lower right). The left set represents a pre-type II burst time, and the right set represents a timestamp during type II at that frequency. The color scale represents the flux density/beam in units of SFU/beam. The cyan circle marks the optical disk of the Sun, and the yellow-filled ellipse in the lower corner of each panel depicts the PSF. The white contours are at 100, 150, 200, 250 $\times\sigma$ for the Stokes I image, and for other Stokes maps, the contours are at $\pm9, \pm21, \pm30$ $\times\sigma$, where $\sigma$ is the rms noise of the corresponding maps. In both sets, positive features are denoted by solid contours and negative features by dashed ones. In the Stokes I map of the pre-burst time (left set), the lower source corresponds to the type II burst, and the upper source corresponds to an active region present during the event.}
    \label{letter_fig:stokes_IQUV_hpc_map}
\end{figure*}

\section{Data Analysis} \label{data-analysis}
\subsection{Calibration and imaging} \label{calibration-and-imaging}
Full-polarization (Stokes I, Q, U, and V) snapshot images of the type II event are produced at a temporal and spectral resolution of 0.5 s and 160 kHz, respectively. Flagging and calibration are performed using the Common Astronomy Software Applications for Radio Astronomy \citep[CASA;][]{casa_Bean2022}, while the imaging was carried out using WSClean \citep{wsclean_offringa2014}. 

The initial bandpass calibration and cross-hand phase (the phase difference between the X and Y dipoles) are carried out using Hercules A (HerA), observed after sunset on the same day at $\sim$10:15:04 UTC, following the methodology presented in \citet{cross_phase_Kansabanik2025}. This corrects for any mixing between Stokes U and V. Mixing between Stokes Q and U, and Stokes Q and V is very small as has been demonstrated in \citet{cross_phase_Kansabanik2025}. Hence, the only remaining polarization mixing is from Stokes I to Stokes Q, U, and V. For self-calibration, we follow the prescription detailed in ``Polarimetry using Automated Imaging Routine for the Compact Arrays for the Radio Sun" \citep[P-AIRCARS;][]{paircars22a,paircars22b,paircars23}.

For an aperture array instrument like the MWA, the primary beam and its polarization characteristics are a strong function of the pointing direction and also vary significantly across its wide field-of-view (FoV).
This direction-dependent primary beam contributes a significant part of the instrumental polarization leakage. 
Developing a detailed and reliable model for the direction-dependent characteristics of the primary beam is hence an essential requirement for achieving high fidelity imaging.
For the MWA, \citet{Sokolowski2017} provide such a Full Embedded Element (FEE) primary beam model.
This model is used to correct for the direction-dependent instrumental leakage.


This model is available at a spectral resolution of 1.28 MHz. 
Our need for higher spectral resolution imaging required us to interpolate the FEE beam to a much finer spectral resolution of 10 kHz. 
The details of the interpolation method of the FEE beam developed as a part of this study are described in Appendix \ref{app:1}. 

Although the FFE model provides a very good representation of the MWA's instrumental beam, some departures from the true instrumental response inevitably arise.
As a consequence, some low-level residual leakage can persist even after applying the correction for the ideal primary beam. This manifests primarily as {\it pol-conversion} \citep{Hamaker2000}, where the Stokes I emission leaks into Stokes Q, U, and V. 

This remaining leakage can vary appreciably across the MWA FoV of several tens of degrees and needs to be accounted for \citep{lenc2017challenges}. However,  \citet{paircars22a} showed that across the solar disk, of angular scale $\sim$ 50-55$\arcmin$ at the observing frequencies, the variation in this leakage is too small to be significant.
Hence, a direction-independent, image-based leakage correction approach is employed, as detailed in \citep{paircars22a}.

Both the ideal primary beam correction and the image-based leakage correction are performed independently at each imaging frequency. After these corrections, the residual leakage across the quiet Sun in the Stokes Q, U, and V maps are found to be noise-like, implying direction-dependent residual leakages, if any, lie below the detection threshold achieved. 
The residual leakages ($L_\mathrm{Q,U,V}$) for Stokes Q, U, and V are computed using the formulation mentioned in \cite{paircars22a} and detailed in Appendix B.

\subsection{Image analysis} \label{subsec:img-analysis}
Some representative full Stokes images (Stokes I, Q, U, and V \citep{stokes1851}) are shown in Figure \ref{letter_fig:stokes_IQUV_hpc_map} as per IAU convention \citep{IAU_1973}. 
The linear polarization intensity maps (P maps) are produced using Stokes Q and U maps as $P = \sqrt{Q^2 + U^2}$. In this work, we primarily focus on the bright sources with more than $9\sigma$ detection in both Stokes I, Q, and U maps, where $\sigma$ represents the rms noise of individual maps. We have then estimated fractional linear polarization and polarization angle maps as follows.

\textbf{Fractional polarization:} 
To estimate the maximum linear polarization fraction, we first identify the location of the peak flux in the $P$ map within the spatial extent of the type II source. A point-spread-function (PSF)-sized region is then defined around this peak location, and the same region is used to measure the average flux densities in the Stokes $Q$, $U$, and $I$ maps, denoted $Q_f$, $U_f$, and $I_f$, respectively. The fractional linear polarization is then estimated as
\begin{equation}
P_f = \sqrt{Q_f^2 + U_f^2},
\end{equation}
\begin{equation}
\mathrm{F_P} = \frac{P_f}{I_f}.
\end{equation}
The uncertainty on $P_f$ follows from standard error propagation,
\begin{equation}
\sigma_{P_f} = \frac{\sqrt{\left(Q_f\,\sigma_{Q_f}\right)^2 + \left(U_f\,\sigma_{U_f}\right)^2}}{P_f},
\end{equation}
where $\sigma_{Q_f}$, $\sigma_{U_f}$ and $\sigma_{I_f}$ are the rms noise of the corresponding Stokes maps.  
The uncertainty on the fractional polarization is
\begin{equation}
\sigma_{\mathrm{F_P}} = \mathrm{F_P}\sqrt{\left(\frac{\sigma_{P_f}}{P_f}\right)^2 + \left(\frac{\sigma_{I_f}}{I_f}\right)^2}.
\end{equation}


A polarized source is detected reliably when both of the following criteria are simultaneously satisfied:
\begin{enumerate}
    \item The linearly polarized signal is detected in the respective polarization (Q, U) maps with more than 9$\sigma$ significance.
    \item The polarization fraction ($\frac{Q}{I}$ or $\frac{U}{I}$) are more that residual leakage limit of the corresponding Stokes maps, $L_\mathrm{Q,U}$, as detailed in Appendix B.
\end{enumerate}

\textbf{Polarization angle:} The polarization angle ($\chi$) is defined as $\chi = \frac{1}{2} \tan^{-1}\left(\frac{U}{Q}\right)$. 
$\chi$ is computed only for regions where both of the criteria given above for regarding the detection of polarized emission as reliable are met.
The uncertainty on $\chi$ is obtained from the rms noise in the Stokes Q and U images using standard error propagation.

\section{Results} \label{sec:results}

Figure \ref{letter_fig:stokes_IQUV_hpc_map} shows two sets of the full-Stokes images -- before (left set) and during (right set) the type II burst at 108.81 MHz.
In the pre-type II Stokes I image, there are two sources with comparatively higher flux density than the quiet Sun (QS), marking the two radio-bright active regions. Neither of these has any associated signature with a significance more than $9\sigma$ in the Stokes Q and U maps. 
The upper Stokes I source is weaker and shows circularly polarized emission, while the lower one is stronger in Stokes I but does not show any associated Stokes V emission. During the type II burst (right set), only the type II source, originating close to the lower active region, is visible in Stokes I, as it is three orders of magnitude brighter than the other active region. The corresponding Stokes Q and U images show compact sources coinciding with the location of the type II burst in the Stokes I image. Each of these sources consists of adjacent emissions of positive and negative polarities, demonstrating varying polarization properties across the spatial extent of the type II source. Both Stokes Q and U signatures have a high signal-to-noise ratio (SNR), indicating the unambiguous presence of linear polarization. The Stokes V image also shows emission with both polarities, with a polarization fraction of $\le 8\%$ for the negative polarity sources. 

Since this work focuses on the observed characteristics of the linearly polarized emission, the rest of this section presents details of spectro-temporal variability of the observed fractional linear polarization ($F_\mathrm{P}$), the source morphology, and the polarization angle ($\chi$).

\begin{figure}[htbp]  
    \centering
    \includegraphics[width=0.95\linewidth]{linear_pol_dynamic_spectrum.png}
    \includegraphics[width=0.93\linewidth]{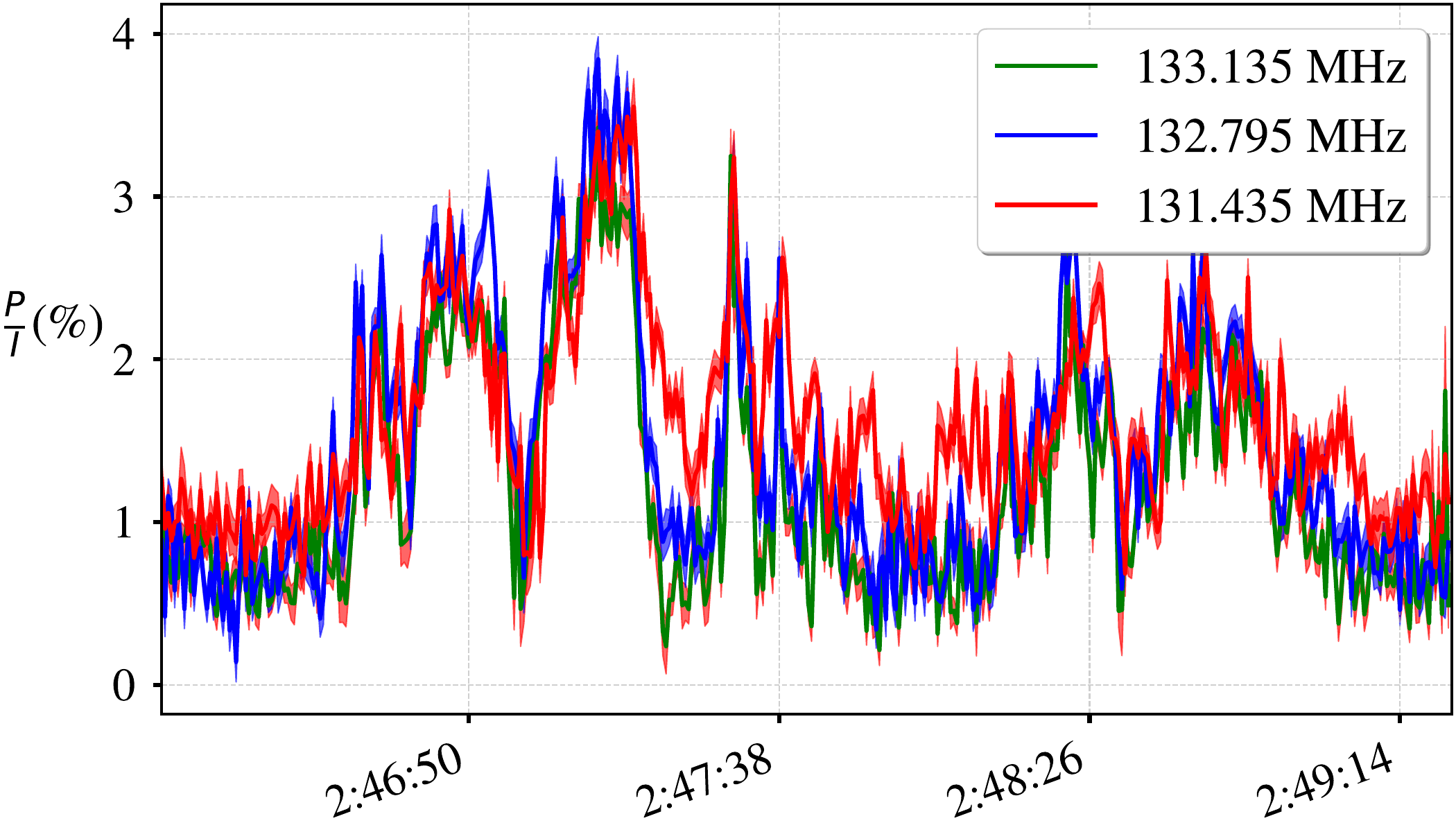}
    \includegraphics[width=0.93\linewidth]{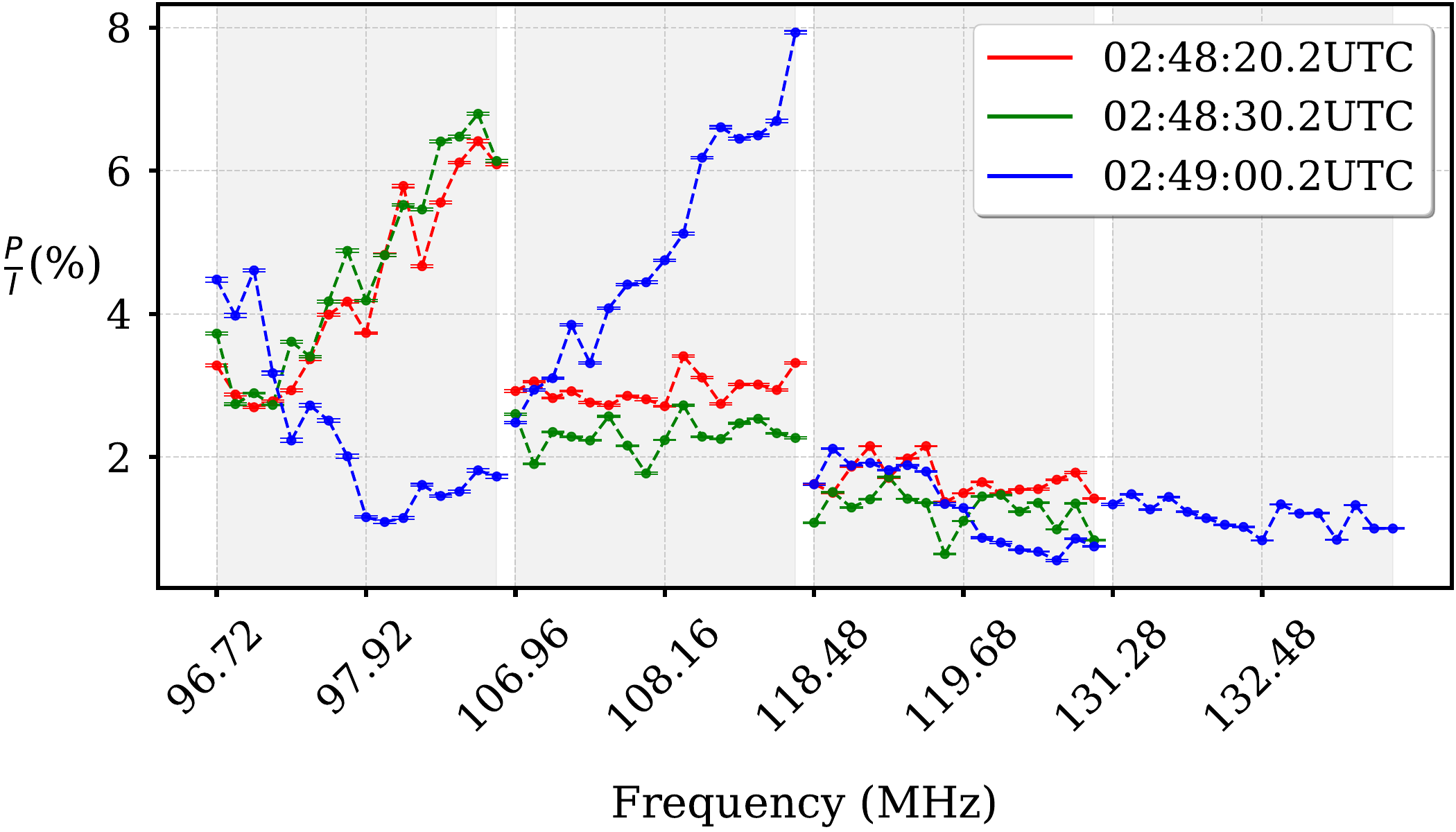}
    \caption{Spectro-temporal variability of $F_\mathrm{P}$ are shown. Top panel: The dynamic spectra of $F_\mathrm{P}$ across six spectral bands are shown. The white regions indicate where the type II burst is absent, as determined by using a threshold on the brightness temperature of the type II source from the Stokes I images.  
    Middle panel: Three traces representing variability of $F_\mathrm{P}$ over time at three nearby frequencies, 131.435 (red), 132.795 (blue), and 133.135 (green) MHz. The shaded regions around each trace represent the associated uncertainties.
    Lower panel: Spectral variability of $F_\mathrm{P}$ at three timestamps 2:48:30 UTC (red), 2:48:30 UTC (blue), 2:49:00 UTC (green), with the associated uncertainties on $F_\mathrm{P}$ shown as error bars. The discontinuous shaded region shows four of the six discrete 2.56 MHz spectral bands in which MWA observed the type-II event at those timestamps.}
    \label{letter_fig:linear_pol_variability}
\end{figure}

\subsection{Spectro-temporal variability of \texorpdfstring{$F_\mathrm{P}$}{FP}} \label{subsec:variability_in_fp}

Figure \ref{letter_fig:linear_pol_variability} shows spectro-temporal variation of $F_\mathrm{P}$, estimated as discussed in Section \ref{calibration-and-imaging}. The upper panel shows the dynamic spectrum of $F_\mathrm{P}$ during the period of the type II burst, while the white color denotes the absence of the burst. $F_\mathrm{P}$ ranges between $\sim$~$3 - 12\%$, exhibiting higher values in the frequency range of 90--108 MHz. We note that despite their seemingly small values in an absolute sense, they have been estimated with a high level of significance and are well above the residual leakages of the corresponding channels (Sec. \ref{calibration-and-imaging}).

The middle panel shows an example of the temporal variation of $F_\mathrm{P}$ at three frequencies, separated by $\sim$340 kHz (green and blue traces) and $\sim$1.36 MHz (red and green traces).
The associated uncertainties on $F_\mathrm{P}$ are shown by shaded regions along with the traces, though these are not very evident as they are too small. Two characteristics of variation in $F_P$ are evident from the figure -- variation ranging over tens of seconds, with successive periods that follow each other very closely at nearby frequencies, with differences increasing slowly as spectral separation increases. Superposed on this longer-term trend are shorter-term fluctuations on time scales as rapid as 0.5 s. These are correlated over narrow spectral separations of $< 1\,\mathrm{MHz}$. We note that, though small, these fluctuations are much larger than the measurement uncertainty.

The bottom plot shows an example of the spectral variation of $F_\mathrm{P}$ at time intervals of 10 s (red and green traces) and 40 s (red and blue traces). 
The spectral evolution of $F_\mathrm{P}$ over a time scale of 10 s is highly correlated, with increasing temporal separation, significant differences in the spectral trends appear.
Note that each of the gray regions represents a discrete 2.56 MHz spectral band and that there are significant spectral gaps between them, as shown by the white rectangles in the left panel of Figure \ref{letter_fig:dynamic_spectra_and_brightness_temperature}.

\subsection{Spectro-temporal variability in source morphology} \label{subsec:variability_in_morphology}
\begin{figure*}[htbp]  
    \centering
    \includegraphics[trim=3.3cm 0.2cm 0.3cm 0.2cm, clip, width=0.32\linewidth]{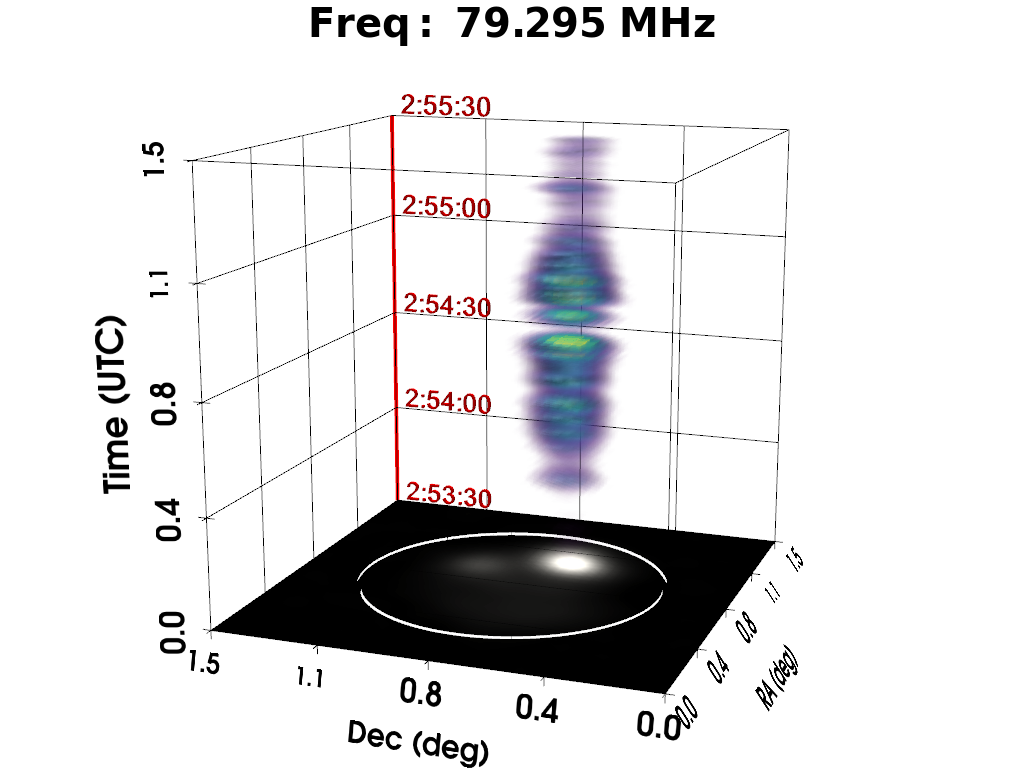}
    \includegraphics[trim=3.3cm 0.2cm 0.3cm 0.2cm, clip, width=0.32\linewidth]{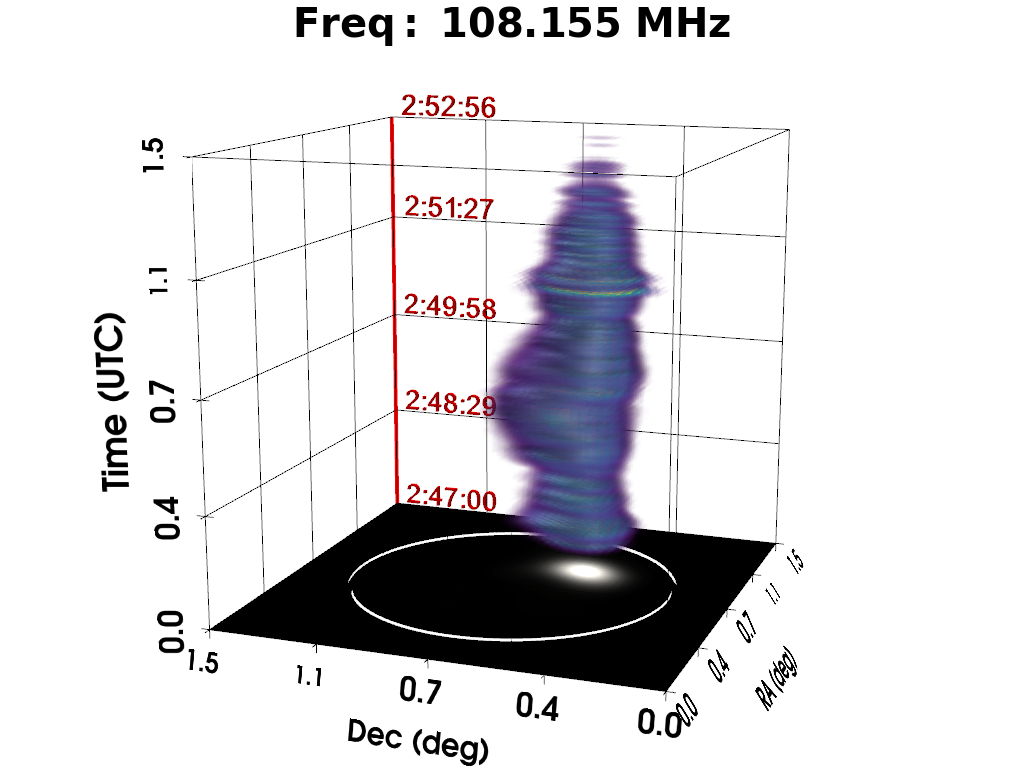}
    \includegraphics[trim=3.3cm 0.2cm 0.3cm 0.2cm, clip, width=0.32\linewidth]{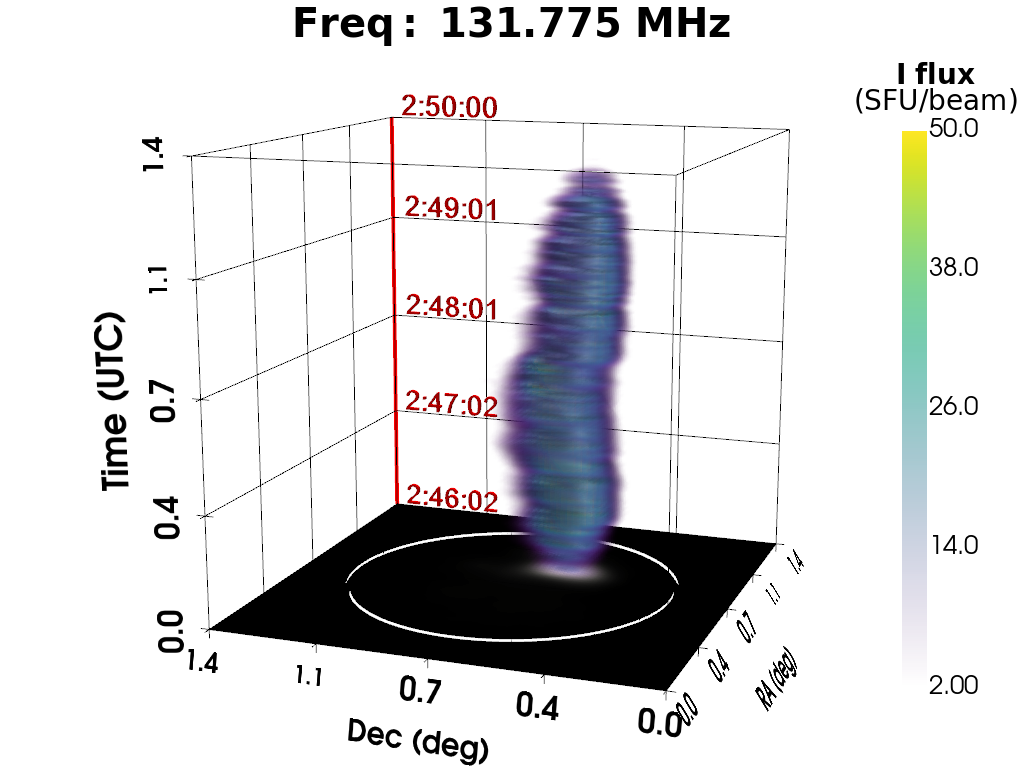}
    \includegraphics[trim=3.3cm 0.2cm 0.3cm 0.2cm, clip, width=0.32\linewidth]{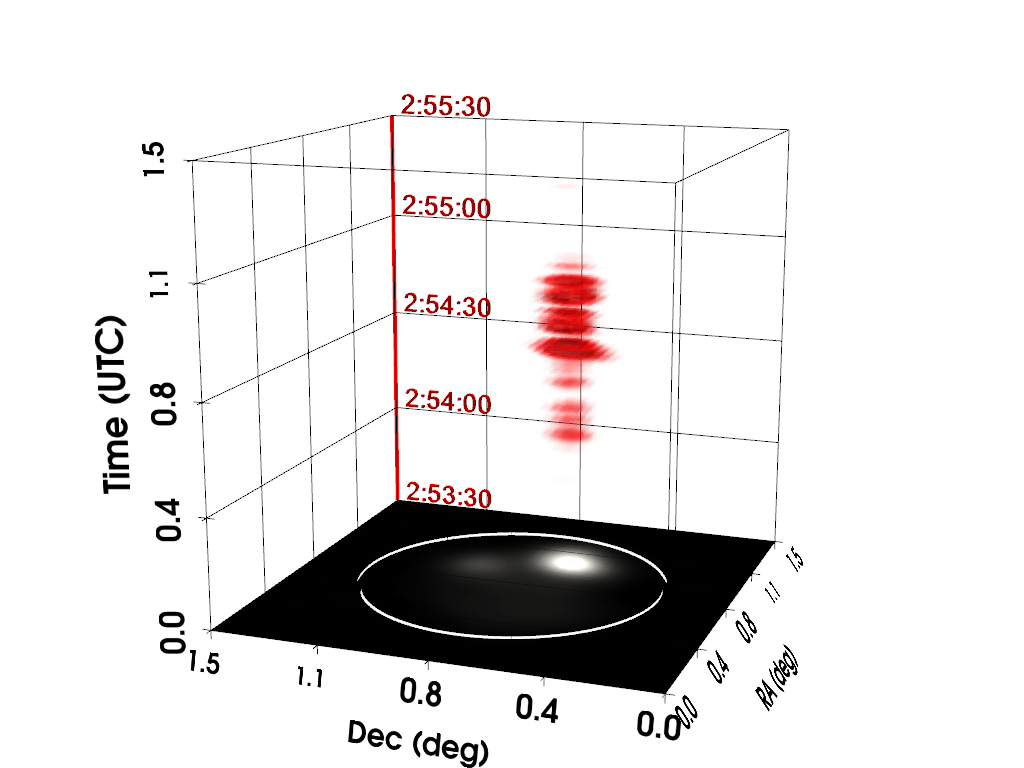}
    \includegraphics[trim=3.3cm 0.2cm 0.3cm 0.2cm, clip, width=0.32\linewidth]{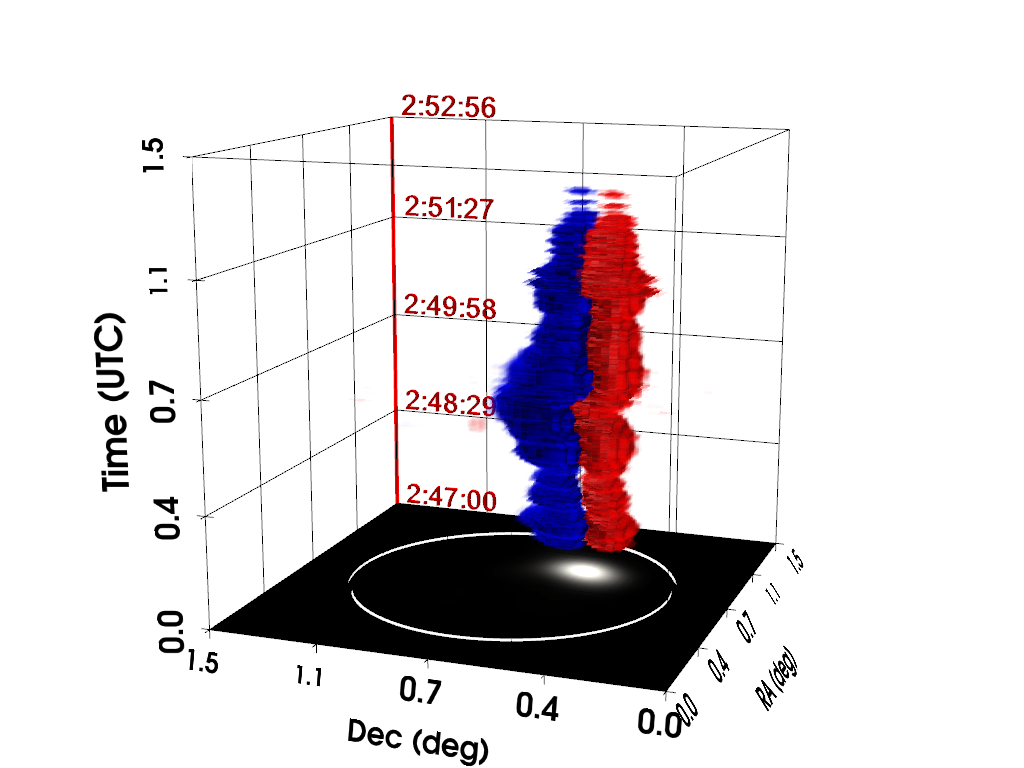}
    \includegraphics[trim=3.3cm 0.2cm 0.3cm 0.2cm, clip, width=0.32\linewidth]{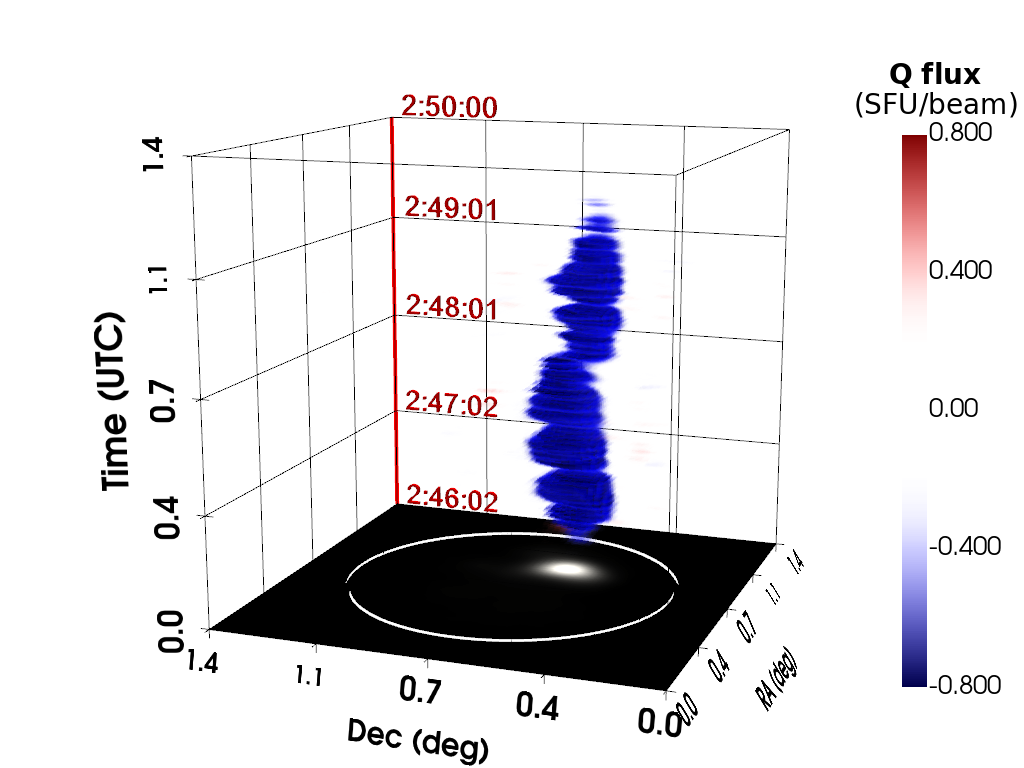}
    \includegraphics[trim=3.3cm 0.2cm 0.3cm 0.2cm, clip, width=0.32\linewidth]{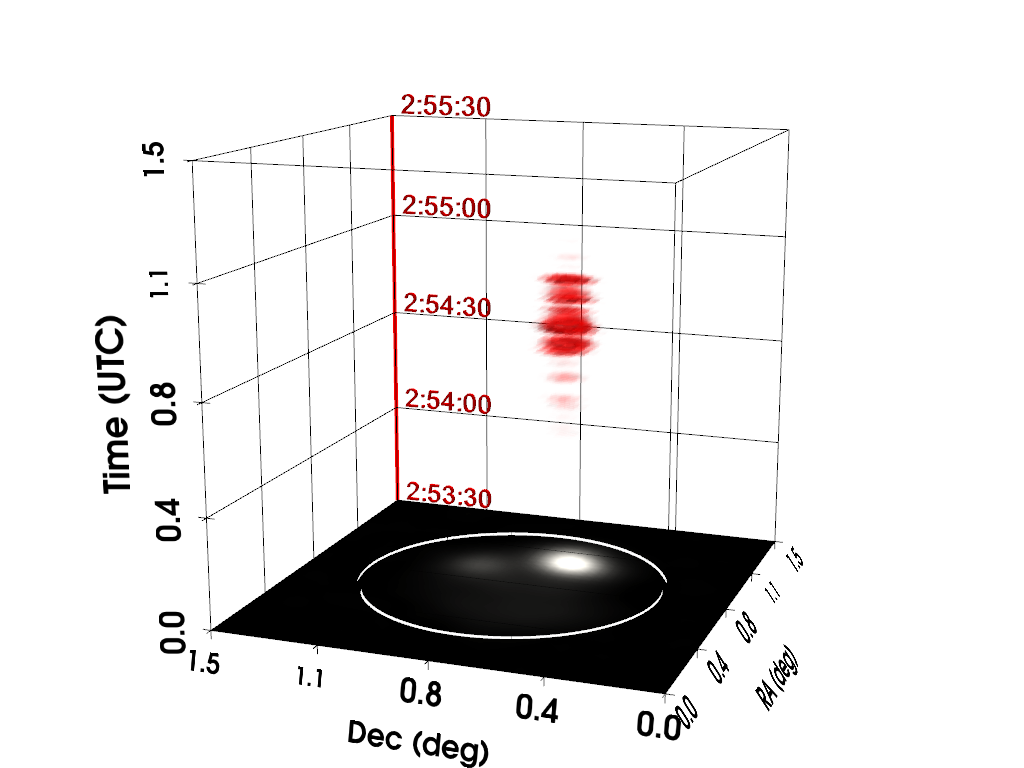}
    \includegraphics[trim=3.3cm 0.2cm 0.3cm 0.2cm, clip, width=0.32\linewidth]{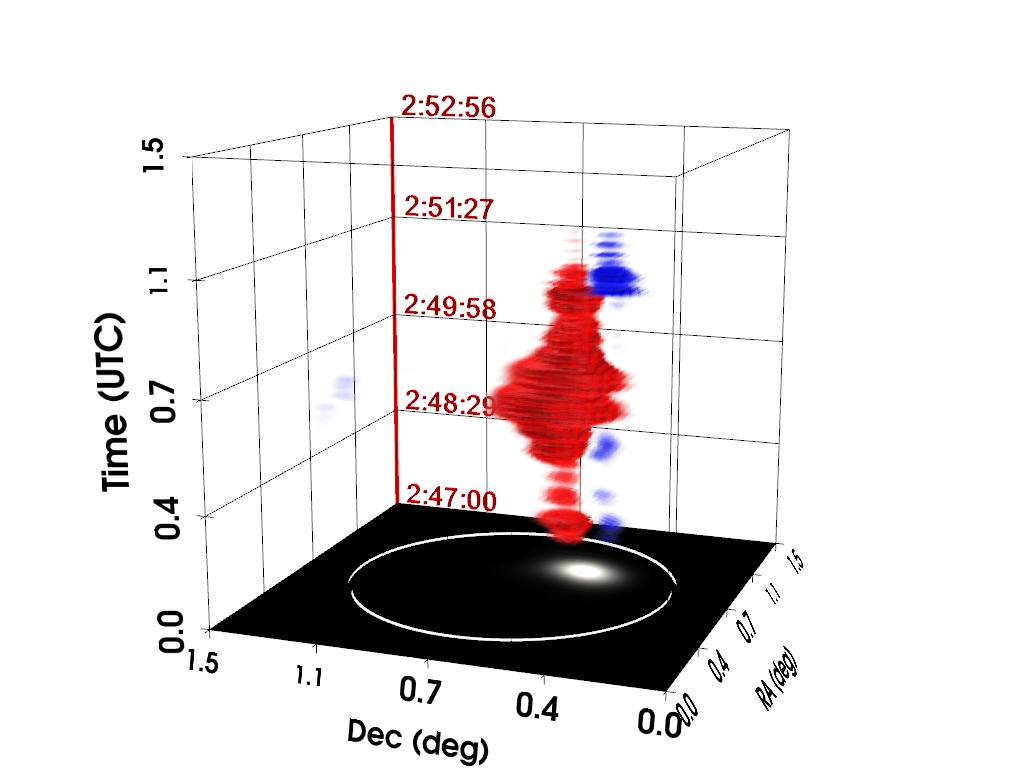}
    \includegraphics[trim=3.3cm 0.2cm 0.3cm 0.2cm, clip, width=0.32\linewidth]{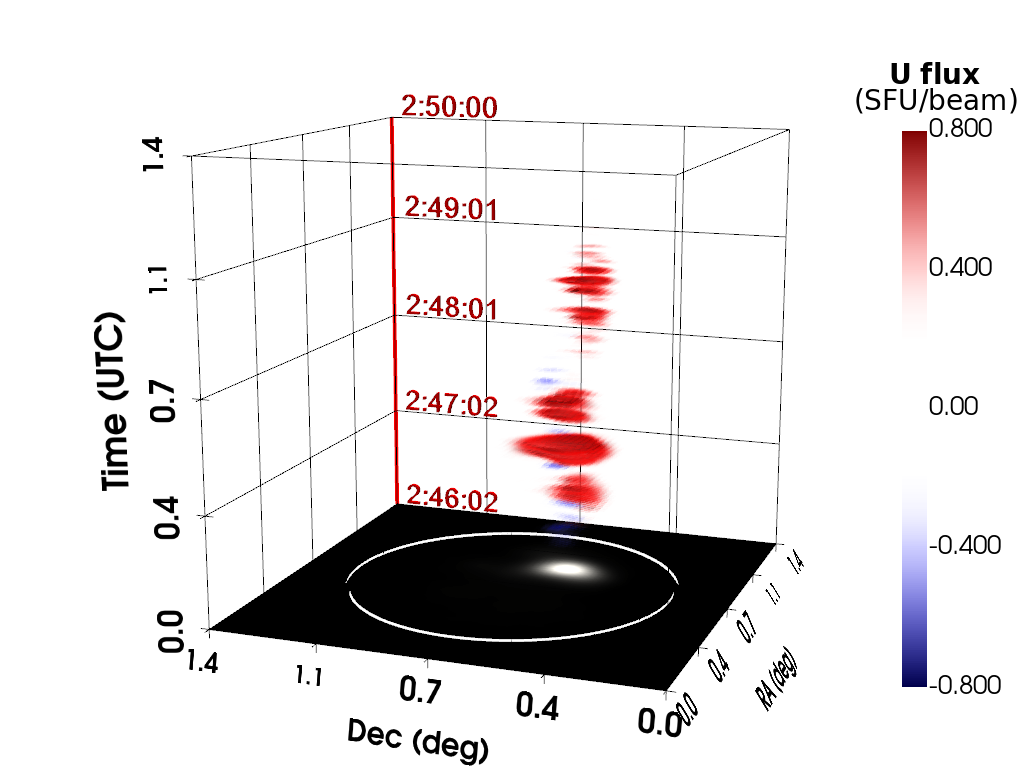}
    \includegraphics[trim=3.3cm 0.2cm 0.3cm 0.2cm, clip, width=0.32\linewidth]{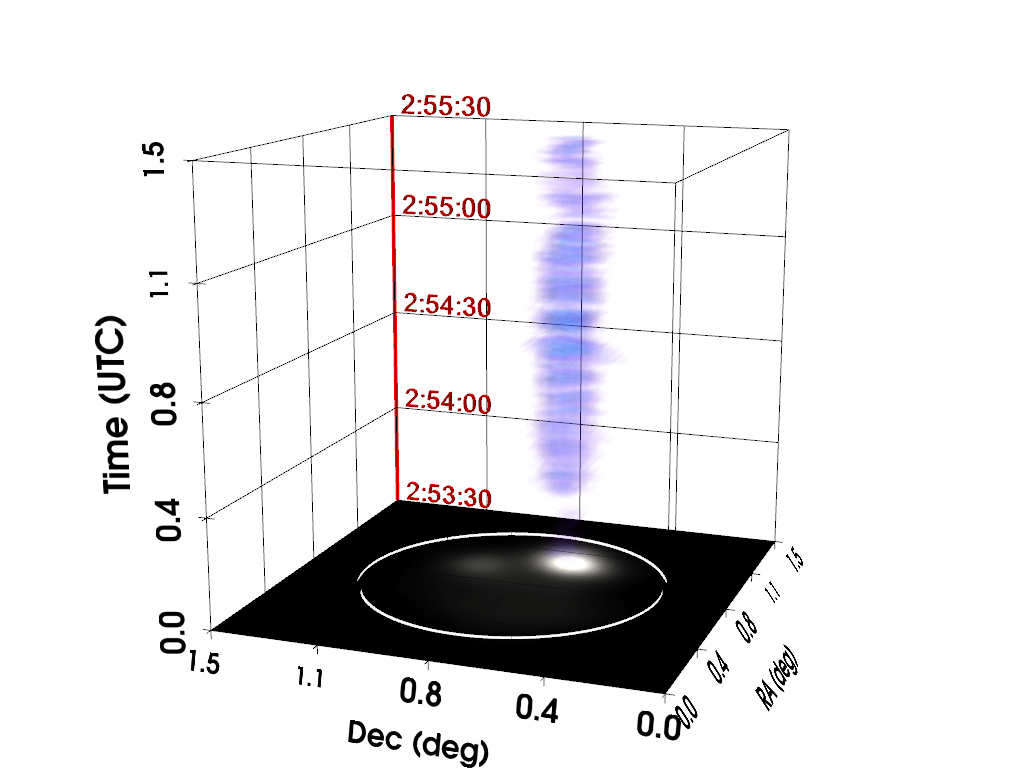}
    \includegraphics[trim=3.3cm 0.2cm 0.3cm 0.2cm, clip, width=0.32\linewidth]{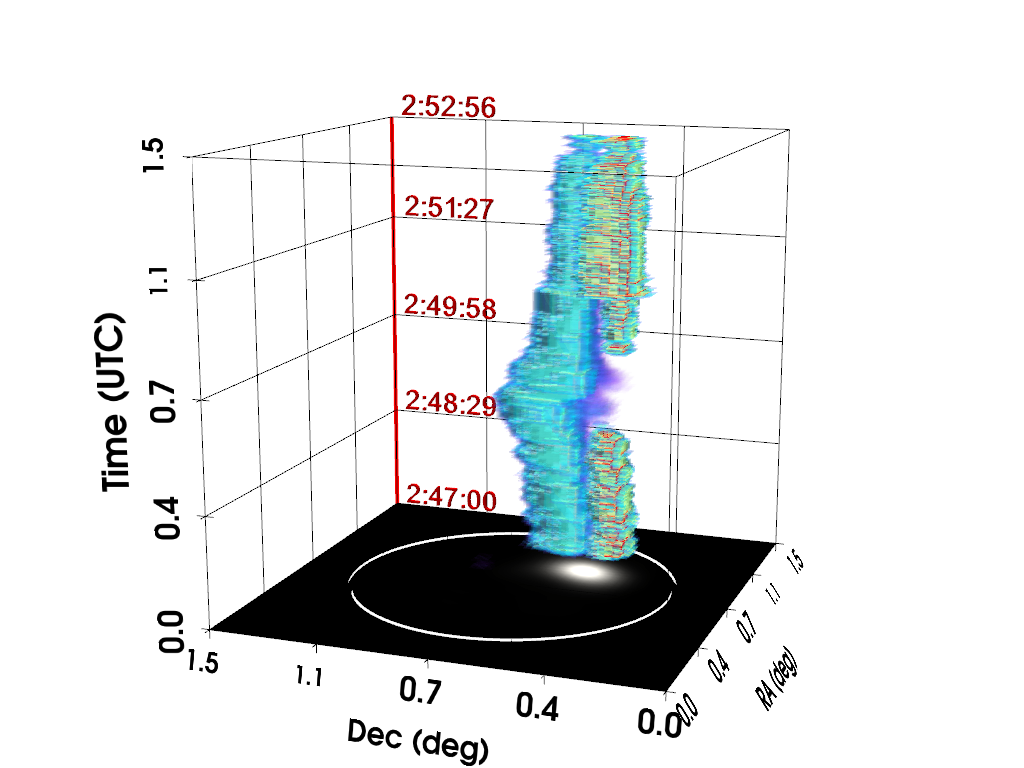}
    \includegraphics[trim=3.3cm 0.2cm 0.3cm 0.2cm, clip, width=0.32\linewidth]{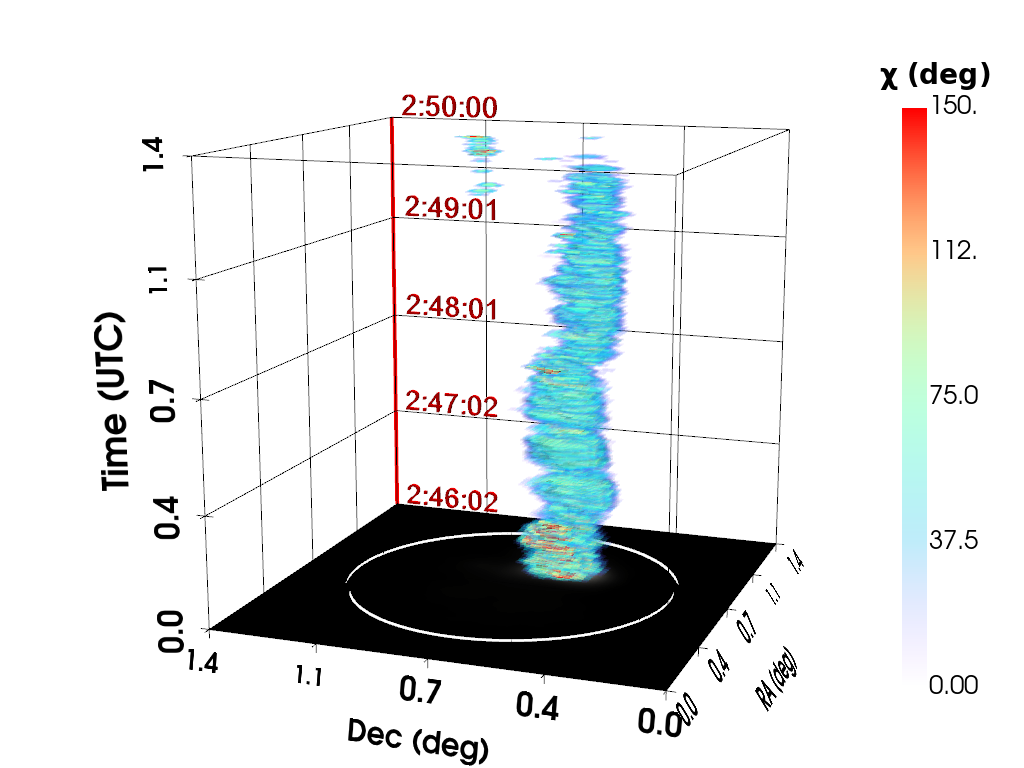}
    \caption{Each panel illustrates the temporal evolution of Stokes I (first row), Q (second row), U (third row) sources, and the polarization angle, $\chi$, (fourth row) at three distinct frequencies, represented by the three columns. The colorbars for the first three rows indicate the flux density in SFU/beam units for Stokes I, Q, and U, respectively, while the colorbar for the fourth row represents $\chi$ in degrees. In each panel, the vertical axis encodes time, and the horizontal plane presents the sky-plane morphology in RA-Dec coordinates in relative units to the origin. The base image in each panel corresponds to the Stokes I image at the onset of the type II burst at the respective frequency, with the white circle denoting the optical solar disk.}
    \label{letter_fig:morphology_temporal_variation}
\end{figure*}

Figure \ref{letter_fig:morphology_temporal_variation} 
presents three-dimensional renderings of the linearly polarized emission from the type II burst at three distinct frequencies -- 79.295 (left column), 108.155 (middle column), and 131.775 (right column) MHz, showing its temporal evolution. Rows from top to bottom show Stokes I, Q, U, and $\chi$, with the vertical axis encoding time and the horizontal axes capturing the sky-plane morphology. 
The variation in Stokes I flux density seen in Figure~\ref{letter_fig:dynamic_spectra_and_brightness_temperature} is also clearly seen in Figure \ref{letter_fig:morphology_temporal_variation}.
In addition to the flux density, the variations in the size and the locations of the Stokes I source are also clearly evident here.
The type II emission is always seen to come from a single Stokes I source. 
At lower frequencies, the centroid of this source remains fairly steady with time, while at higher frequencies, e.g., 108.155 and 131.775 MHz, it shows an abrupt shift of the centroid at around 02:49:00 UTC and 2:48:02 UTC, respectively. 
This shift is more evident in the temporal evolution of the Stokes Q source centroids.

Figure \ref{letter_fig:polarization_angle_variation_morphology} shows contours of Stokes Q (white) and U (cyan) at all six frequencies for a few selected time stamps spanning the duration of the type II burst at that frequency. Sources with positive and negative polarities are depicted by solid and dashed contours, respectively.
Stokes Q sources show a trend in polarity across frequency. At the highest frequency, the Stokes Q shows emission with only negative polarity. With decreasing frequency, the emission first becomes bipolar, and then at the lowest frequency, it shows only positive polarity.
At a given frequency, the nature of Stokes Q polarity remains mostly consistent throughout the event, as can be seen in Figure \ref{letter_fig:morphology_temporal_variation}. 
Stokes U images show a higher level of variability across both time and frequency.
Figure~\ref{letter_fig:morphology_temporal_variation} shows the broad trend that the polarities of the observed Stokes U source change systematically from being positive at the highest frequency, to becoming bipolar at the mid frequencies, and becoming positive again at the lowest frequencies. 
There are also deviations from these trends, at least for some of the times and frequencies (Figure~\ref{letter_fig:polarization_angle_variation_morphology}).

\subsection{Spectro-temporal variability in \texorpdfstring{$\chi$}{chi}} \label{subsec:variability_in_pol_ang}
\begin{figure*}[htbp]  
    \centering
    \includegraphics[trim=0.5cm 0.5cm 0.5cm 0.5cm, clip, width=1.02\linewidth]{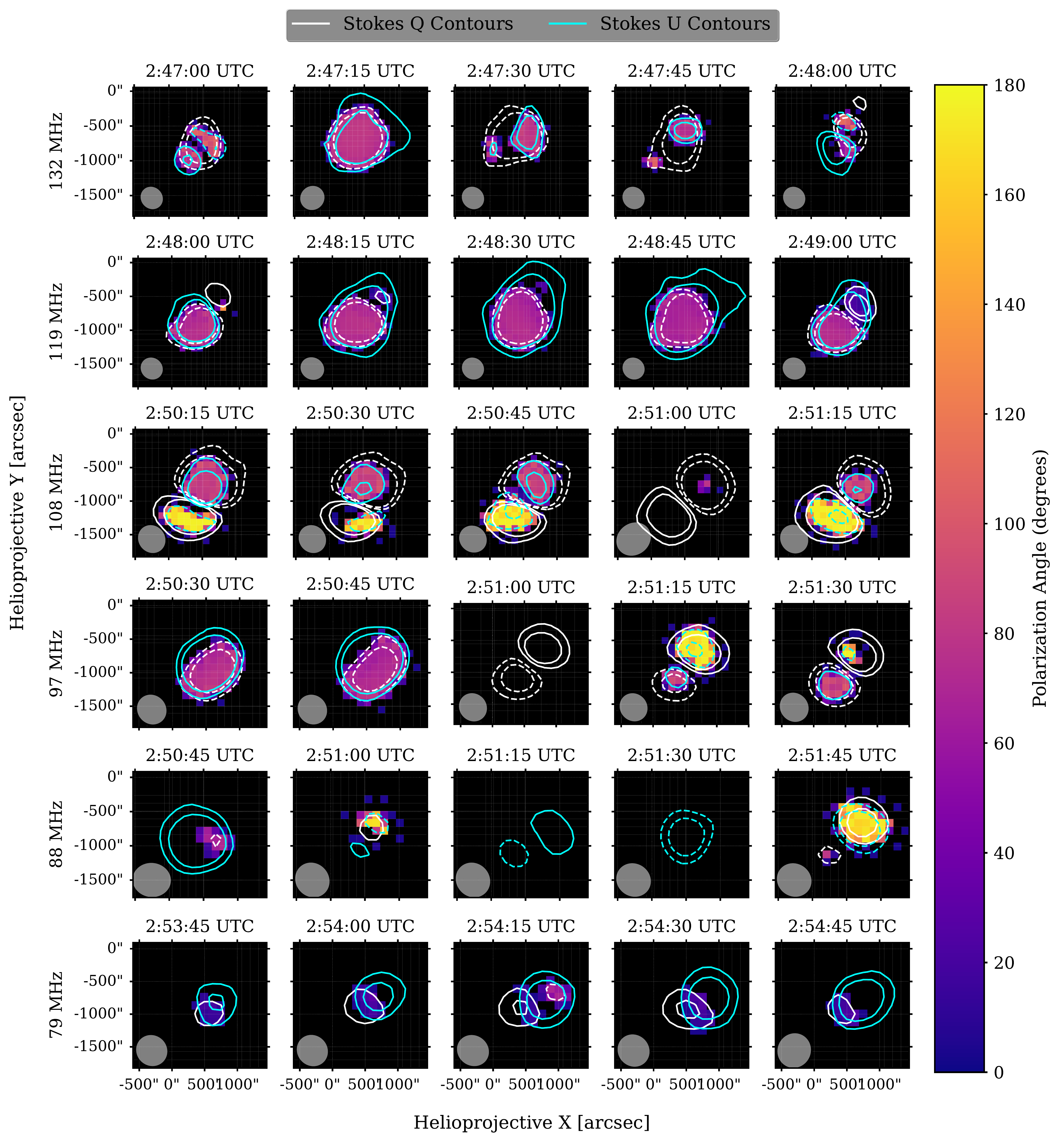}
    \caption{Polarization angle maps at the type II burst source location, where panels of each row correspond to polarization maps in helioprojective coordinates at different timestamps of a single frequency channel. The colorbar presents the polarization angles in degrees. The grey ellipse in the lower-left corner of each panel represents the PSF of the image. The white and cyan contours correspond to the 10, 20 $\times \sigma$ (where $\sigma$ is the rms noise of the map) levels of the Stokes Q and U maps, respectively, where the solid and dashed contours represent the positive and the negative levels.}
    \label{letter_fig:polarization_angle_variation_morphology}
\end{figure*}

The bottom panel of Figure \ref{letter_fig:morphology_temporal_variation} shows the temporal variation of $\chi$ throughout the duration of the burst, at three distinct frequencies, and Figure 
\ref{letter_fig:polarization_angle_variation_morphology} shows the spatial variability of $\chi$ across the observed spectral coverage of the burst at a few timestamps. $\chi$ is determined as described in Section \ref{subsec:img-analysis}.
As is evident, $\chi$ is determined with a high level of confidence.
$\chi$ shows very little spatial variability at any single time and frequency, while the source size is much broader than the PSF (Figure \ref{letter_fig:polarization_angle_variation_morphology}).
For any given frequency, it remains stable across time for the entire duration of the type II burst (bottom panel of Figure \ref{letter_fig:morphology_temporal_variation}).
When two linearly polarized sources are simultaneously present, as seen in 108 MHz, their $\chi$ values are quite different ($\sim82^\circ$ and $\sim174^\circ$), but remain stable across the burst duration. 
Though $\chi$ does change across frequency, as evident in Figure \ref{letter_fig:polarization_angle_variation_morphology}, it does not seem to follow an obvious trend.

\section{Discussion} \label{sec:discussion}
It has long been believed that, even if linear polarization were to be intrinsically present, the high Faraday rotation (FR) encountered during coronal propagation will reduce any linear polarization to zero \citep{Grognard1973}.
As an illustration - considering the hybrid electron density model proposed by \citep{Vrsnak2004} and assuming the fundamental emission to be at $\sim$ 65 MHz, the type II burst is estimated to originate at a heliocentric distance of 1.49 $R_\odot$. Adopting the coronal magnetic model by \citet{Dulk_McLean_1978} leads to a rotation measure (RM) of $\sim$ 2.6 $\times 10^5$ $rad/m^2$ for this coronal height. 
This corresponds to a differential FR ($\Delta\psi$) -- the difference in rotation angle accumulated across a finite observing bandwidth 
leading to depolarization of the intrinsic linear polarization due to averaging across the bandwidth --
of $\sim$7400 rad across a bandwidth of just 160 kHz used in this study, at the central observing frequency of 100 MHz. It is estimated as $\Delta\psi = \text{RM} \times 2\lambda\,\Delta\lambda$
, where $\lambda = c/\nu$ is the central wavelength and $\Delta\lambda = c\,\Delta\nu/\nu^2$ is the wavelength spread corresponding to the bandwidth $\Delta\nu$, yielding $\Delta\psi \gg 2\pi$ -- far exceeding the threshold for depolarization of any linearly polarized emission when averaged over even a narrow bandwidth.

However, \citet{Dey_2025_linear} have recently demonstrated the first surprising but robust imaging detection of linearly polarized metric solar emission from types I and III solar radio bursts. A key part of their argument was demonstrating consistent measurements of linearly polarized emission in simultaneous observations at the same frequency using two very different telescopes, the MWA and upgraded Giant Metrewave Radio Telescope \citep[uGMRT;][]{uGMRT2017}, which are geographically separated by thousands of km.

\subsection{Reliability of detected linearly polarized emission}
\label{sub_sec:reliable_detection}

We present next the arguments for why the linearly polarized emission presented in Section \ref{sec:results} cannot be of instrumental origin.
The MWA, like all aperture array instruments, does indeed have large instrumental leakage. Its simple design and lack of moving parts, however, ensure that this leakage varies slowly, smoothly, and systematically with time, frequency, and pointing direction, and can also be modeled in detail \citep{lenc2017challenges, Sokolowski2017, cross_phase_Kansabanik2025}.
As argued in Sec \ref{calibration-and-imaging}, any residual leakage across the solar disc is not expected to be direction-dependent.
If the observed linearly polarized emission were to arise due to instrumental leakage, its fractional polarization should remain constant across the solar disc, and its morphology should closely resemble that seen in Stokes I images, neither of which is observed to be true. Evidence for these arguments can be realized from Figures \ref{letter_fig:linear_pol_variability} and \ref{letter_fig:stokes_IQUV_hpc_map}.
Figure \ref{letter_fig:linear_pol_variability} shows 
rapid temporal and spectral variations in the observed linear polarization fraction, which cannot arise due to any instrumental leakages. 
Figure \ref{letter_fig:stokes_IQUV_hpc_map} shows the Stokes Q, U, and V images in the pre-burst time (left set). These pre-burst images contain two bright compact emissions in Stokes I images, arising from presence of active regions. Except for these two bright Stokes I  regions, rest of the solar disk emission comes from QS. None of the Stokes Q, U or V images show any resemblance to the QS emission seen in the Stokes I image.

Figures \ref{letter_fig:morphology_temporal_variation} and \ref{letter_fig:polarization_angle_variation_morphology} show multiple instances where the observed linearly polarized emission morphology changes remarkably across a few arcminutes. For instance, the row corresponding to 97 MHz in Figure \ref{letter_fig:polarization_angle_variation_morphology} shows a transition from a single negative Q and positive U lobes to adjacent positive and negative Q and U lobes. 
Similarly, in the row corresponding to 88 MHz, the observed U emission goes from being only positive to adjacent positive and negative lobes to only negative.
The panel corresponding to 108 MHz, Stokes U emission in Figure \ref{letter_fig:morphology_temporal_variation} shows that the emission is dominantly positive, though a negative source also appears for some instances. This is also not consistent with the expectations from instrumental leakage, which changes slowly and smoothly across the FoV and also remains stable across a time span of 4 min.
All of these arguments imply that the detected linear polarization must indeed be present in the radiation incident on the telescope and cannot arise from any calibration or instrumental artifacts.

\subsection{Origin of linear polarization} \label{subsec:origin_of_lpoli}
We have already established that the observed linear polarization is not of instrumental origin.
Additionally, Figures \ref{letter_fig:morphology_temporal_variation} and \ref{letter_fig:polarization_angle_variation_morphology} show that a given patch of linearly polarized emission maintains a rather narrow distribution of $\chi$ across all the observed frequencies. The value of $\chi$ remains stable throughout the duration of the burst at any given frequency, yet differs from one frequency band to another. Note that type II emission at different frequencies originates from spatially distinct regions as the shock front traverses the coronal density gradient.
Each frequency, hence, traces a different path through the intervening corona. The stability of $\chi$ throughout the event at each frequency implies that once generated, the linearly polarized emission is essentially unaffected by FR. 

The linear polarization detected from the harmonic type II emission must either be intrinsic to the burst or arise as a consequence of the propagation effects along the ray paths.
As already discussed earlier in this section, under usual circumstances, the linear polarization generated in the vicinity of the emission site is expected to be washed out due to the enormous coronal FR. 
The only way for it to survive is if there exist conditions for the coronal depolarization to be much less effective. 
Recently, independent evidence for a discrepancy between the observed size of type I sources and the expectations based on well-regarded coronal scattering models has emerged, with observations reporting source sizes $\sim$3 times smaller than the predictions \citep[][Dey et al., in prep]{Mondal2024, Mondal2025}. 
A possible way suggested in those studies to resolve the discrepancy is by embedding the radio source in an overdense region at larger coronal heights as compared to the expectations based on coronal electron density models \citep{Mondal2025}. This ensures that, despite the emission being at fundamental and/or harmonic frequencies, it emerges from the compact overdense region into a much rarer medium and hence experiences a much lower FR.
Using the hybrid electron density model by \citet{Vrsnak2004} and the coronal magnetic field model by \citet{Dulk_McLean_1978}, at an observing frequency of 100 MHz, the differential FR across 160 kHz drops below $< 1$ radian for a line of sight (LoS) passing at the heliocentric distance of $\sim$ 7.7 $R_\odot$. This implies that, in order to be consistent with the negligible observed FR, the linear polarization must arise beyond the coronal height of $\sim$ 7.7 $R_\odot$.

\FloatBarrier

\citet{Bastian2022} suggested reflection from an overdense region as a possible mechanism for the generation of linear polarization to explain their observation of UV Ceti.
The generation of linear polarization is more effective at incident angles approaching the Brewster's angle. To reduce the coronal FR post-reflection, this overdense region must lie at much higher coronal heights.
The observed temporal evolution of Stokes Q and U sources and the juxtaposition of positive and negative Stokes Q sources, however, argue against the idea of a common reflection geometry.

Another intriguing mechanism for the generation of linearly polarized emission is through mode-coupling during quasi-transverse (QT) propagation, when the wave propagation direction is nearly perpendicular to the background magnetic field \citep{cohen1960}. In this geometry, the characteristic polarization of the two magnetoionic modes -- extraordinary (x) and ordinary (o) modes -- changes rapidly along the ray path, creating conditions under which the two modes can exchange energy, a phenomenon known as mode coupling. 
Metric type II bursts arise at a lower coronal height, which has a magnetically complex and dynamic environment. This naturally presents numerous opportunities for the emission to encounter QT propagation along the ray path. The degree to which the two modes couple in a QT region is quantified by a dimensionless coupling parameter $C$ \citep{cohen1960}, which depends on the observing frequency, local plasma conditions, i.e., electron density and strength of the background magnetic field, and the scale length over which the magnetic field orientation changes. In the weak coupling regime ($C\ll 1$), the two modes propagate independently and reconcile their polarization depending upon the relative orientation of the background magnetic field. In the strong coupling regime ($C\gg 1$), the modes are tightly coupled and thus the various magnetoionic phenomena, such as FR and differential absorptions, are ineffective. In the transition regime ($C\sim 1$), partial energy transfer between the x and o modes occurs, generating a linear polarization component from an initially circularly polarized emission.

While simple mode coupling theory predicts weak coupling in QT regions at metre wavelengths \citep{cohen1960}, several observational findings suggest that the coupling is in fact moderate to strong at these wavelengths. These include the absence of polarization sense reversal near meridian passage in type I bursts \citep{Piddington1951}, the anomalously low degree of circular polarization in the fundamental emission of type II and III bursts \citep{Zheleznyakov1964}, and the recent proposal of mode coupling as a generation mechanism for linear polarization in metric solar bursts \citep{Dey_2025_linear}. The coupling strength can be enhanced by reducing the characteristic length scale over which the magnetic field orientation changes, thereby increasing $C$ toward the moderate-to-strong regime. Several physical mechanisms have been proposed to achieve this at metre wavelengths, including magnetic field rippling due to Alfv\'{e}n wave propagation \citep{melrose_1974}, the presence of coronal current sheets \citep{Gopalswamy1994}, and angular scattering off coronal turbulence \citep{Bastian1995} — all of which can drive the coupling into the moderate-to-strong regime even at metre wavelengths \citep{Bastian1995}.

This study also finds mode coupling during QT propagation to be the physically most plausible mechanism responsible for the generation of the observed linear polarization.
We hypothesize that as the initially weakly circularly polarized harmonic emission propagates outward through the corona, it encounters a QT region in the transitional regime ($C\sim 1$), where partial mode conversion generates a linear polarization component. Beyond this region, the emission enters the strong coupling regime ($C\gg 1$), where the polarization state is frozen and insensitive to further magnetic field variations, consistent with the observed temporal stability of the polarization angle $\chi$ across all frequency bands. 
We also note the linear polarization is always low -- within the 3--12$\%$ regime -- and varies across frequency, suggesting that some frequencies encounter more favorable coupling conditions than others, pointing to a structured and non-uniform magnetic field topology along the propagation path or perhaps the degree of depolarization is different for different frequencies.

\section{Conclusion} \label{sec:conclusion}
This study presents the first detection of linear polarization in a metric type II solar radio burst. 
This type II burst was associated with a CME, and the harmonic emission was identified and imaged across six spectral bands spanning 78--133 MHz.
It exhibits weak circular polarization ($< 8\%$), broadly consistent with theoretical expectations for harmonic plasma emission, and a linear polarization fraction of up to $12\%$ measured with a high level of significance.

The Stokes Q and U sources display more elaborate structure and temporal variability than the Stokes I source, pointing to the presence of variations in the local condition in the medium through which the radiation propagates. The variability in source morphology and the $F_\mathrm{P}$ at close-by frequencies exhibit a correlated trend, with differences increasing with increasing spectral separation. 
This suggests that with increasing spectral separation, the emission also travels through increasingly separated regions with different local plasma environments.
The linearly polarized emission is confined to a narrow range of $\chi$ across the observed frequencies, and the angle remains stable throughout the duration of the type II emission. 
Mode-coupling during QT propagation is the most likely mechanism for the generation of linearly polarized emission. 

The presence of linearly polarized emission has now been conclusively shown in types I, II, and III solar radio bursts.  
Its presence across different types of bursts implies that it is a generic feature of most types of active solar emissions. Their ubiquity also suggests that it might arise due to a propagation effect, irrespective of the details of the physical conditions at the emission site and emission process.

This work yields an entirely new observable for investigating and understanding coronal conditions and propagation. 
It also points to gaps and limitations in our current understanding of these areas.
This was made possible by the high-quality spectro-polarimetric snapshot radio imaging capability offered by the recent generation of radio interferometers, which enabled the exploration of a previously inaccessible part of phase space. 
With the Square Kilometre Array Observatory expected to be operational early next decade, this phase space is expected to grow very substantially.
We hope that these results and the promise of even better imaging, which will enable detailed tests of any proposed hypotheses, will spur the community to take up suitable investigations to improve our understanding in these areas.

\section{Acknowledgement}
This scientific work uses data obtained from Inyarrimanha Ilgari Bundara, the CSIRO Murchison Radio-astronomy Observatory. We acknowledge the Wajarri Yamaji People as the Traditional Owners and Native Title Holders of the observatory site. Support for the operation of the MWA is provided by the Australian Government (NCRIS), under a contract to Curtin University administered by Astronomy Australia Limited. 
We thank Dr. Surajit Mondal for his valuable comments and helpful discussions.
P.M. and D.O. acknowledge the support of the Department of Atomic Energy, Government of India, under project no. 12-R\&D-TFR-5.02-0700. D.K. acknowledges financial support from the Severo Ochoa grant CEX2021-001131-S and from the Spanish grant PID2023-147883NB-C21, funded by MCIU/AEI/ 10.13039/501100011033, as well as support through ERDF/EU, and the NASA Living with a Star Jack Eddy Postdoctoral Fellowship Program, administered by UCAR’s Cooperative Programs for the Advancement of Earth System Science (CPAESS) under award 80NSSC22M0097.
The authors gratefully acknowledges the developers of CASA \citep{casa_Bean2022}, SunPy \citep{sunpy_community2020}, astropy \citep{astropy2013, astropy2018, astropy2022}, matplotlib \citep{Hunter:2007}, Numpy \citep{harris2020array}, PyVista \citep{sullivan2019pyvista}.

\appendix

\section{Spectral Interpolation of MWA Primary Beam Model}\label{app:1}
The most advanced primary beam model currently available for the MWA is the Full Embedded Element (FEE) model described by \cite{Sokolowski2017}. In this model, the primary beam response of an MWA antenna tile, comprising 16 bow-tie dipoles arranged in a \(4 \times 4\) configuration, is estimated through full electromagnetic simulations. The beam response is represented using a set of spherical harmonic coefficients (\(H\)) corresponding to each antenna element in the tile (see Equation~6 of \cite{Sokolowski2017}). Since calculating these coefficients through full electromagnetic simulations is computationally very expensive, \cite{Sokolowski2017} generated them only for the discrete 1.28~MHz coarse channels of the MWA. Consequently, beam estimation at an arbitrary frequency is typically obtained by selecting the nearest available coarse-channel model, introducing low-level systematic effects in polarization measurements. To mitigate these effects, we implemented a frequency interpolation of the primary beam model instead of relying on the nearest-frequency approximation.

For this interpolation, we chose to interpolate the \(H\) coefficients directly because they are independent of the beamformer pointing direction. This approach enables the same interpolated coefficients to be used for any beamformer pointing without requiring separate interpolations. The coefficients are expressed as
\[
H = H_\mathrm{real} + jH_\mathrm{imag},
\]
where \(H_\mathrm{real}\) and \(H_\mathrm{imag}\) denote the real and imaginary components, respectively. Cubic spline interpolation was applied independently to both \(H_\mathrm{real}\) and \(H_\mathrm{imag}\) as functions of frequency using the values available at 1.28~MHz intervals. The interpolated functions were then used to estimate \(H\) coefficients at the desired finer spectral resolution (\(<1.28\)~MHz). These interpolated coefficients were subsequently incorporated into the same formalism described in Equations~3--5 of \cite{Sokolowski2017}. The interpolated FEE primary beam model adopted in this work significantly reduces the low-level polarization systematics as compared to the FEE beam models modeled at a spectral resolution of 1.28~MHz.

\section{Estimation of Residual leakages}\label{app:2}
\begin{figure*}[htbp]  
    \centering
    \includegraphics[width=0.49\linewidth]{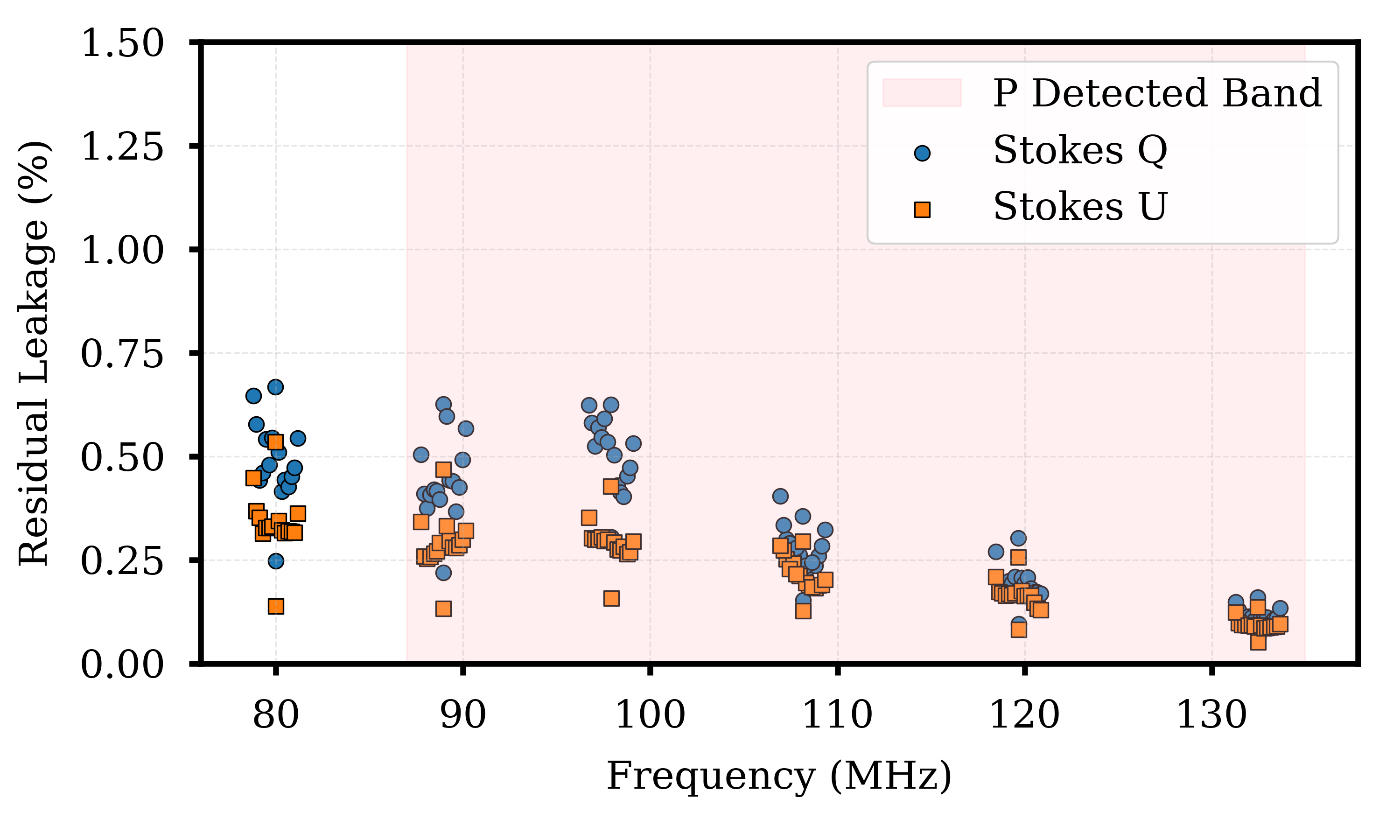}
    \includegraphics[width=0.49\linewidth]{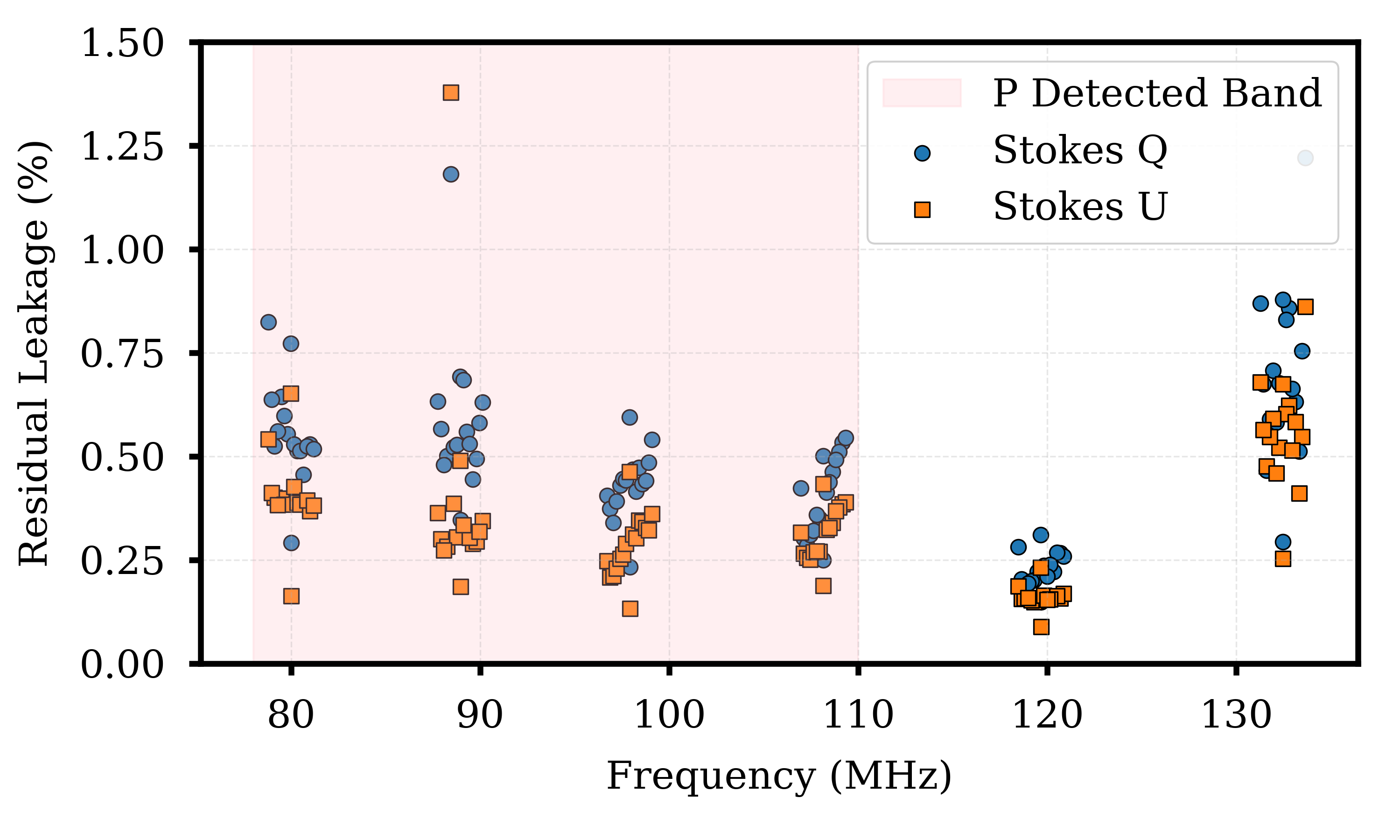}
    \caption{Residual leakage as a function of frequency across the six observed spectral bands, shown separately for the two consecutive datasets. Blue circles mark the residual leakage in the Stokes Q images, and red squares mark that in the Stokes U images. The pink shaded regions in the background indicate the spectral bands over which the type II burst was detected in each respective dataset.}
    \label{letter_fig:residual_leakages_across_datasets}
\end{figure*} 
Following the ideal primary beam correction and image-based leakage correction, the residual leakage across the quiet Sun in the Stokes Q and U maps is found to be consistent with noise.
To estimate this residual leakage, we apply lower and upper brightness temperature thresholds, derived from the Stokes I map, to mask the active region and the region outside the solar disc. This ensures that only quiet-Sun emission is included in the Stokes Q and U maps. The residual leakage across the quiet Sun is then quantified using a $3\sigma$ criterion, which provides an upper limit on the residual leakage as follows:\\
\begin{equation}
L_{\mathrm{Q,U}} =
\left|\frac{3 \times \sigma_{Q,U}}{I_{\mathrm{max}}}\right|,
\end{equation}
where $\sigma_{Q,U}$ denotes the rms noise of the Stokes Q or U image, and $I_{\mathrm{max}}$ is the peak Stokes I flux density within the quiet-Sun region. The resulting values across all six spectral bands are shown in Figure \ref{letter_fig:residual_leakages_across_datasets}. 
As noted in sec \ref{subsec:img-analysis}, calibration and imaging are performed independently for the two consecutive 4-minute datasets, and the panels of Figure \ref{letter_fig:residual_leakages_across_datasets} show the residual leakage obtained for both of them.

\bibliography{sample631}{}

@ARTICLE{uGMRT2017,
       author = {{Gupta}, Y. and {Ajithkumar}, B. and {Kale}, H.~S. and {Nayak}, S. and {Sabhapathy}, S. and {Sureshkumar}, S. and {Swami}, R.~V. and {Chengalur}, J.~N. and {Ghosh}, S.~K. and {Ishwara-Chandra}, C.~H. and {Joshi}, B.~C. and {Kanekar}, N. and {Lal}, D.~V. and {Roy}, S.},
        title = "{The upgraded GMRT: opening new windows on the radio Universe}",
      journal = {Current Science},
         year = 2017,
        month = aug,
       volume = {113},
       number = {4},
        pages = {707-714},
          doi = {10.18520/cs/v113/i04/707-714},
       adsurl = {https://ui.adsabs.harvard.edu/abs/2017CSci..113..707G}
}

@article{Hamaker2000,
	author = {{Hamaker, J. P.}},
	title = {Understanding radio polarimetry - IV. The full-coherency analogue of scalar self-calibration:
	Self-alignment, dynamic range and polarimetric fidelity},
	DOI= "10.1051/aas:2000337",
	url= "https://doi.org/10.1051/aas:2000337",
	journal = {Astron. Astrophys. Suppl. Ser.},
	year = 2000,
	volume = 143,
	number = 3,
	pages = "515-534",
}

@Article{IAU_1973,
title={Commission 40: Radio Astronomy (Radio Astronomie)},
journal={Transactions of the International Astronomical Union},
author={IAU},
year={1973},
edition={2018/03/13},
publisher={Cambridge University Press},
volume={15},
number={2},
pages={165-167},
issn={0251-107X},
doi={10.1017/S0251107X00031606},
url={https://www.cambridge.org/core/article/commission-40-radio-astronomy-radio-astronomie/0A120377080E784F27148CFAD0049564}
}

@ARTICLE{stokes1851,
       author = {{Stokes}, G.~G.},
        title = "{On the Composition and Resolution of Streams of Polarized Light from different Sources}",
      journal = {Transactions of the Cambridge Philosophical Society},
         year = 1851,
        month = jan,
       volume = {9},
        pages = {399},
       adsurl = {https://ui.adsabs.harvard.edu/abs/1851TCaPS...9..399S}
}

@article{roberts1959solar,
  title={Solar radio bursts of spectral type II},
  author={Roberts, JA},
  journal={Australian Journal of Physics},
  volume={12},
  number={4},
  pages={327--356},
  year={1959},
  publisher={CSIRO Publishing}
}

@ARTICLE{Sokolowski2017,
       author = {{Sokolowski}, M. and {Colegate}, T. and {Sutinjo}, A.~T. and {Ung}, D. and {Wayth}, R. and {Hurley-Walker}, N. and {Lenc}, E. and {Pindor}, B. and {Morgan}, J. and {Kaplan}, D.~L. and {Bell}, M.~E. and {Callingham}, J.~R. and {Dwarakanath}, K.~S. and {For}, Bi-Qing and {Gaensler}, B.~M. and {Hancock}, P.~J. and {Hindson}, L. and {Johnston-Hollitt}, M. and {Kapi{\'n}ska}, A.~D. and {McKinley}, B. and {Offringa}, A.~R. and {Procopio}, P. and {Staveley-Smith}, L. and {Wu}, C. and {Zheng}, Q.},
        title = "{Calibration and Stokes Imaging with Full Embedded Element Primary Beam Model for the Murchison Widefield Array}",
      journal = {\pasa},
         year = 2017,
        month = nov,
       volume = {34},
          eid = {e062},
        pages = {e062},
          doi = {10.1017/pasa.2017.54},
archivePrefix = {arXiv},
       eprint = {1710.07478},
 primaryClass = {astro-ph.IM},
       adsurl = {https://ui.adsabs.harvard.edu/abs/2017PASA...34...62S}
}

@ARTICLE{wsclean_offringa2014,
       author = {{Offringa}, A.~R. and {McKinley}, B. and {Hurley-Walker}, N. and {Briggs}, F.~H. and {Wayth}, R.~B. and {Kaplan}, D.~L. and {Bell}, M.~E. and {Feng}, L. and {Neben}, A.~R. and {Hughes}, J.~D. and {Rhee}, J. and {Murphy}, T. and {Bhat}, N.~D.~R. and {Bernardi}, G. and {Bowman}, J.~D. and {Cappallo}, R.~J. and {Corey}, B.~E. and {Deshpande}, A.~A. and {Emrich}, D. and {Ewall-Wice}, A. and {Gaensler}, B.~M. and {Goeke}, R. and {Greenhill}, L.~J. and {Hazelton}, B.~J. and {Hindson}, L. and {Johnston-Hollitt}, M. and {Jacobs}, D.~C. and {Kasper}, J.~C. and {Kratzenberg}, E. and {Lenc}, E. and {Lonsdale}, C.~J. and {Lynch}, M.~J. and {McWhirter}, S.~R. and {Mitchell}, D.~A. and {Morales}, M.~F. and {Morgan}, E. and {Kudryavtseva}, N. and {Oberoi}, D. and {Ord}, S.~M. and {Pindor}, B. and {Procopio}, P. and {Prabu}, T. and {Riding}, J. and {Roshi}, D.~A. and {Shankar}, N. Udaya and {Srivani}, K.~S. and {Subrahmanyan}, R. and {Tingay}, S.~J. and {Waterson}, M. and {Webster}, R.~L. and {Whitney}, A.~R. and {Williams}, A. and {Williams}, C.~L.},
        title = "{WSCLEAN: an implementation of a fast, generic wide-field imager for radio astronomy}",
      journal = {\mnras},
         year = 2014,
        month = oct,
       volume = {444},
       number = {1},
        pages = {606-619},
          doi = {10.1093/mnras/stu1368},
archivePrefix = {arXiv},
       eprint = {1407.1943},
 primaryClass = {astro-ph.IM},
       adsurl = {https://ui.adsabs.harvard.edu/abs/2014MNRAS.444..606O}
}

@ARTICLE{casa_Bean2022,
       author = {{CASA Team} and {Bean}, Ben and {Bhatnagar}, Sanjay and {Castro}, Sandra and {Donovan Meyer}, Jennifer and {Emonts}, Bjorn and {Garcia}, Enrique and {Garwood}, Robert and {Golap}, Kumar and {Gonzalez Villalba}, Justo and {Harris}, Pamela and {Hayashi}, Yohei and {Hoskins}, Josh and {Hsieh}, Mingyu and {Jagannathan}, Preshanth and {Kawasaki}, Wataru and {Keimpema}, Aard and {Kettenis}, Mark and {Lopez}, Jorge and {Marvil}, Joshua and {Masters}, Joseph and {McNichols}, Andrew and {Mehringer}, David and {Miel}, Renaud and {Moellenbrock}, George and {Montesino}, Federico and {Nakazato}, Takeshi and {Ott}, Juergen and {Petry}, Dirk and {Pokorny}, Martin and {Raba}, Ryan and {Rau}, Urvashi and {Schiebel}, Darrell and {Schweighart}, Neal and {Sekhar}, Srikrishna and {Shimada}, Kazuhiko and {Small}, Des and {Steeb}, Jan-Willem and {Sugimoto}, Kanako and {Suoranta}, Ville and {Tsutsumi}, Takahiro and {van Bemmel}, Ilse M. and {Verkouter}, Marjolein and {Wells}, Akeem and {Xiong}, Wei and {Szomoru}, Arpad and {Griffith}, Morgan and {Glendenning}, Brian and {Kern}, Jeff},
        title = "{CASA, the Common Astronomy Software Applications for Radio Astronomy}",
      journal = {\pasp},
         year = 2022,
        month = nov,
       volume = {134},
       number = {1041},
          eid = {114501},
        pages = {114501},
          doi = {10.1088/1538-3873/ac9642},
archivePrefix = {arXiv},
       eprint = {2210.02276},
 primaryClass = {astro-ph.IM},
       adsurl = {https://ui.adsabs.harvard.edu/abs/2022PASP..134k4501C}
}

@ARTICLE{Alissandrakis1994_linear_pol,
       author = {{Alissandrakis}, C.~E. and {Chiuderi-Drago}, F.},
        title = "{Detection of Linear Polarization in the Microwave Emission of Solar Active Regions}",
      journal = {\apjl},
         year = 1994,
        month = jun,
       volume = {428},
        pages = {L73},
          doi = {10.1086/187396},
       adsurl = {https://ui.adsabs.harvard.edu/abs/1994ApJ...428L..73A}
}

@article{Bhunia_2023,
   title={Imaging-spectroscopy of a band-split type II solar radio burst with the Murchison Widefield Array},
   volume={670},
   ISSN={1432-0746},
   url={http://dx.doi.org/10.1051/0004-6361/202244456},
   DOI={10.1051/0004-6361/202244456},
   journal={Astronomy \&amp; Astrophysics},
   publisher={EDP Sciences},
   author={Bhunia, Shilpi and Carley, Eoin P. and Oberoi, Divya and Gallagher, Peter T.},
   year={2023},
   month=feb, pages={A169} }

@ARTICLE{Vrsnak2002,
       author = {{Vr{\v{s}}nak}, B. and {Magdaleni{\'c}}, J. and {Aurass}, H. and {Mann}, G.},
        title = "{Band-splitting of coronal and interplanetary type II bursts. II. Coronal magnetic field and Alfv{\'e}n velocity}",
      journal = {\aap},
         year = 2002,
        month = dec,
       volume = {396},
        pages = {673-682},
          doi = {10.1051/0004-6361:20021413},
       adsurl = {https://ui.adsabs.harvard.edu/abs/2002A\&A...396..673V}
}

@ARTICLE{Vrsnak2001,
       author = {{Vr{\v{s}}nak}, B. and {Aurass}, H. and {Magdaleni{\'c}}, J. and {Gopalswamy}, N.},
        title = "{Band-splitting of coronal and interplanetary type II bursts. I. Basic properties}",
      journal = {\aap},
         year = 2001,
        month = oct,
       volume = {377},
        pages = {321-329},
          doi = {10.1051/0004-6361:20011067},
       adsurl = {https://ui.adsabs.harvard.edu/abs/2001A\&A...377..321V}
}

@ARTICLE{Vrsnak2004,
       author = {{Vr{\v{s}}nak}, B. and {Magdaleni{\'c}}, J. and {Zlobec}, P.},
        title = "{Band-splitting of coronal and interplanetary type II bursts. III. Physical conditions in the upper corona and interplanetary space}",
      journal = {\aap},
         year = 2004,
        month = jan,
       volume = {413},
        pages = {753-763},
          doi = {10.1051/0004-6361:20034060},
       adsurl = {https://ui.adsabs.harvard.edu/abs/2004A\&A...413..753V}
}

@ARTICLE{Kondo1995,
       author = {{Kondo}, Tetsuro and {Isobe}, Takeshi and {Igi}, Seiji and {Watari}, Shinichi and {Tokumaru}, Munetoshi},
        title = "{The Hiraiso Radio Spectrograph (HiRAS) for monitoring solar radio bursts.}",
      journal = {Journal of the Communications Research Laboratory},
         year = 1995,
        month = may,
       volume = {42},
        pages = {111},
       adsurl = {https://ui.adsabs.harvard.edu/abs/1995CLRJ...42..111K}
}

@ARTICLE{paircars22a,
       author = {{Kansabanik}, Devojyoti and {Oberoi}, Divya and {Mondal}, Surajit},
        title = "{Tackling the Unique Challenges of Low-frequency Solar Polarimetry with the Square Kilometre Array Low Precursor: The Algorithm}",
      journal = {\apj},
         year = 2022,
        month = jun,
       volume = {932},
       number = {2},
          eid = {110},
        pages = {110},
          doi = {10.3847/1538-4357/ac6758},
archivePrefix = {arXiv},
       eprint = {2204.04578},
 primaryClass = {astro-ph.SR},
       adsurl = {https://ui.adsabs.harvard.edu/abs/2022ApJ...932..110K}
}

@ARTICLE{paircars22b,
       author = {Kansabanik, Devojyoti and {Oberoi}, Divya and {Mondal}, Surajit},
        title = "{A novel algorithm for high fidelity spectro-polarimetric snapshot imaging of the low-frequency radio Sun using SKA-low precursor}",
      journal = {arXiv e-prints},
         year = 2022,
        month = jul,
          eid = {arXiv:2207.11924},
        pages = {arXiv:2207.11924},
          doi = {10.48550/arXiv.2207.11924},
archivePrefix = {arXiv},
       eprint = {2207.11924},
 primaryClass = {astro-ph.SR},
       adsurl = {https://ui.adsabs.harvard.edu/abs/2022arXiv220711924K}
}

@ARTICLE{paircars23,
       author = {{Kansabanik}, Devojyoti and {Bera}, Apurba and {Oberoi}, Divya and {Mondal}, Surajit},
        title = "{Tackling the Unique Challenges of Low-frequency Solar Polarimetry with the Square Kilometre Array Low Precursor: Pipeline Implementation}",
      journal = {\apjs},
         year = 2023,
        month = feb,
       volume = {264},
       number = {2},
          eid = {47},
        pages = {47},
          doi = {10.3847/1538-4365/acac79},
archivePrefix = {arXiv},
       eprint = {2209.06666},
 primaryClass = {astro-ph.SR},
       adsurl = {https://ui.adsabs.harvard.edu/abs/2023ApJS..264...47K}
}

@ARTICLE{Zheleznyakov1964,
       author = {{Zheleznyakov}, V.~V. and {Zlotnik}, E. Ya.},
        title = "{Polarization of Radio Waves Passing through a Transverse Magnetic Field Region in the Solar Corona}",
      journal = {\sovast},
         year = 1964,
        month = feb,
       volume = {7},
        pages = {485},
       adsurl = {https://ui.adsabs.harvard.edu/abs/1964SvA.....7..485Z}
}

@article{lenc2017challenges,
  title={The challenges of low-frequency radio polarimetry: Lessons from the Murchison Widefield Array},
  author={Lenc, Emil and Anderson, CS and Barry, N and Bowman, JD and Cairns, IH and Farnes, JS and Gaensler, BM and Heald, G and Johnston-Hollitt, M and Kaplan, DL and others},
  journal={Publications of the Astronomical Society of Australia},
  volume={34},
  pages={e040},
  year={2017},
  publisher={Cambridge University Press}
}

@ARTICLE{Hariharan2014,
       author = {{Hariharan}, K. and {Ramesh}, R. and {Kishore}, P. and {Kathiravan}, C. and {Gopalswamy}, N.},
        title = "{An Estimate of the Coronal Magnetic Field near a Solar Coronal Mass Ejection from Low-frequency Radio Observations}",
      journal = {\apj},
         year = 2014,
        month = nov,
       volume = {795},
       number = {1},
          eid = {14},
        pages = {14},
          doi = {10.1088/0004-637X/795/1/14},
       adsurl = {https://ui.adsabs.harvard.edu/abs/2014ApJ...795...14H}
}

@ARTICLE{Smerd1975,
       author = {{Smerd}, S.~F. and {Sheridan}, K.~V. and {Stewart}, R.~T.},
        title = "{Split-band structure in type II radio bursts from the sun}",
      journal = {\aplett},
         year = 1975,
        month = jan,
       volume = {16},
       number = {1},
        pages = {23-28},
       adsurl = {https://ui.adsabs.harvard.edu/abs/1975ApL....16R..23S}
}

@ARTICLE{Tingay2013_mwa,
       author = {{Tingay}, S.~J. and {Goeke}, R. and {Bowman}, J.~D. and {Emrich}, D. and {Ord}, S.~M. and {Mitchell}, D.~A. and {Morales}, M.~F. and {Booler}, T. and {Crosse}, B. and {Wayth}, R.~B. and {Lonsdale}, C.~J. and {Tremblay}, S. and {Pallot}, D. and {Colegate}, T. and {Wicenec}, A. and {Kudryavtseva}, N. and {Arcus}, W. and {Barnes}, D. and {Bernardi}, G. and {Briggs}, F. and {Burns}, S. and {Bunton}, J.~D. and {Cappallo}, R.~J. and {Corey}, B.~E. and {Deshpande}, A. and {Desouza}, L. and {Gaensler}, B.~M. and {Greenhill}, L.~J. and {Hall}, P.~J. and {Hazelton}, B.~J. and {Herne}, D. and {Hewitt}, J.~N. and {Johnston-Hollitt}, M. and {Kaplan}, D.~L. and {Kasper}, J.~C. and {Kincaid}, B.~B. and {Koenig}, R. and {Kratzenberg}, E. and {Lynch}, M.~J. and {Mckinley}, B. and {Mcwhirter}, S.~R. and {Morgan}, E. and {Oberoi}, D. and {Pathikulangara}, J. and {Prabu}, T. and {Remillard}, R.~A. and {Rogers}, A.~E.~E. and {Roshi}, A. and {Salah}, J.~E. and {Sault}, R.~J. and {Udaya-Shankar}, N. and {Schlagenhaufer}, F. and {Srivani}, K.~S. and {Stevens}, J. and {Subrahmanyan}, R. and {Waterson}, M. and {Webster}, R.~L. and {Whitney}, A.~R. and {Williams}, A. and {Williams}, C.~L. and {Wyithe}, J.~S.~B.},
        title = "{The Murchison Widefield Array: The Square Kilometre Array Precursor at Low Radio Frequencies}",
      journal = {\pasa},
         year = 2013,
        month = jan,
       volume = {30},
          eid = {e007},
        pages = {e007},
          doi = {10.1017/pasa.2012.007},
archivePrefix = {arXiv},
       eprint = {1206.6945},
 primaryClass = {astro-ph.IM},
       adsurl = {https://ui.adsabs.harvard.edu/abs/2013PASA...30....7T}
}

@ARTICLE{Wayth2018_mwa,
       author = {{Wayth}, Randall B. and {Tingay}, Steven J. and {Trott}, Cathryn M. and {Emrich}, David and {Johnston-Hollitt}, Melanie and {McKinley}, Ben and {Gaensler}, B.~M. and {Beardsley}, A.~P. and {Booler}, T. and {Crosse}, B. and {Franzen}, T.~M.~O. and {Horsley}, L. and {Kaplan}, D.~L. and {Kenney}, D. and {Morales}, M.~F. and {Pallot}, D. and {Sleap}, G. and {Steele}, K. and {Walker}, M. and {Williams}, A. and {Wu}, C. and {Cairns}, Iver. H. and {Filipovic}, M.~D. and {Johnston}, S. and {Murphy}, T. and {Quinn}, P. and {Staveley-Smith}, L. and {Webster}, R. and {Wyithe}, J.~S.~B.},
        title = "{The Phase II Murchison Widefield Array: Design overview}",
      journal = {\pasa},
         year = 2018,
        month = nov,
       volume = {35},
          eid = {e033},
        pages = {e033},
          doi = {10.1017/pasa.2018.37},
archivePrefix = {arXiv},
       eprint = {1809.06466},
 primaryClass = {astro-ph.IM},
       adsurl = {https://ui.adsabs.harvard.edu/abs/2018PASA...35...33W}
}

@ARTICLE{cross_phase_Kansabanik2025,
       author = {{Kansabanik}, Devojyoti and {Vourlidas}, Angelos and {Dey}, Soham and {Mondal}, Surajit and {Oberoi}, Divya},
        title = "{A Formalism for Calibrating the Instrumental Polarization of Radio Interferometric Arrays at Meter Wavelengths Using Unpolarized Sky: A Demonstration Using MWA Observations}",
      journal = {\apjs},
         year = 2025,
        month = may,
       volume = {278},
       number = {1},
          eid = {26},
        pages = {26},
          doi = {10.3847/1538-4365/adc443},
archivePrefix = {arXiv},
       eprint = {2503.16675},
 primaryClass = {astro-ph.IM},
       adsurl = {https://ui.adsabs.harvard.edu/abs/2025ApJS..278...26K}
}

@ARTICLE{Du2014,
       author = {{Du}, Guohui and {Chen}, Yao and {Lv}, Maoshui and {Kong}, Xiangliang and {Feng}, Shiwei and {Guo}, Fan and {Li}, Gang},
        title = "{Temporal Spectral Shift and Polarization of a Band-splitting Solar Type II Radio Burst}",
      journal = {\apjl},
         year = 2014,
        month = oct,
       volume = {793},
       number = {2},
          eid = {L39},
        pages = {L39},
          doi = {10.1088/2041-8205/793/2/L39},
       adsurl = {https://ui.adsabs.harvard.edu/abs/2014ApJ...793L..39D}
}

@article{Ramesh_2023,
 adsurl = {https://ui.adsabs.harvard.edu/abs/2023ApJ...943...43R},
 archiveprefix = {arXiv},
 author = {{Ramesh}, R. and {Kathiravan}, C. and {Kumari}, Anshu},
 doi = {10.3847/1538-4357/acaea5},
 eid = {43},
 eprint = {2302.00071},
 journal = {\apj},
 month = {January},
 number = {1},
 pages = {43},
 primaryclass = {astro-ph.SR},
 title = {{Solar Coronal Density Turbulence and Magnetic Field Strength at the Source Regions of Two Successive Metric Type II Radio Bursts}},
 volume = {943},
 year = {2023}
}

@article{Ramesh_and_Kathiravan2022,
 adsurl = {https://ui.adsabs.harvard.edu/abs/2022ApJ...940...80R},
 author = {{Ramesh}, R. and {Kathiravan}, C.},
 doi = {10.3847/1538-4357/ac9c65},
 eid = {80},
 journal = {\apj},
 month = {November},
 number = {1},
 pages = {80},
 title = {{Polarization Observations of a Split-band Type II Radio Burst from the Solar Corona}},
 volume = {940},
 year = {2022}
}

@ARTICLE{Grognard1973,
       author = {{Grognard}, R.~J.~M. and {McLean}, D.~J.},
        title = "{Non-Existence of Linear Polarization in Type III Solar Bursts at 80 MHz}",
      journal = {\solphys},
         year = 1973,
        month = mar,
       volume = {29},
       number = {1},
        pages = {149-161},
          doi = {10.1007/BF00153446},
       adsurl = {https://ui.adsabs.harvard.edu/abs/1973SoPh...29..149G}
}

@ARTICLE{Nindos2011,
       author = {{Nindos}, A. and {Alissandrakis}, C.~E. and {Hillaris}, A. and {Preka-Papadema}, P.},
        title = "{On the relationship of shock waves to flares and coronal mass ejections}",
      journal = {\aap},
         year = 2011,
        month = jul,
       volume = {531},
          eid = {A31},
        pages = {A31},
          doi = {10.1051/0004-6361/201116799},
archivePrefix = {arXiv},
       eprint = {1105.1268},
 primaryClass = {astro-ph.SR},
       adsurl = {https://ui.adsabs.harvard.edu/abs/2011A&A...531A..31N}
}

@ARTICLE{Melrose_1989,
       author = {Melrose, D.~B.},
        title = "{Depolarization of Solar Bursts due to Scattering by Low-Frequency Waves}",
      journal = {\solphys},
         year = 1989,
        month = mar,
       volume = {119},
       number = {1},
        pages = {143-156},
          doi = {10.1007/BF00146218},
       adsurl = {https://ui.adsabs.harvard.edu/abs/1989SoPh..119..143M}
}

@ARTICLE{melrose_1974,
       author = {{Melrose}, D.~B.},
        title = "{Mode coupling in the solar corona. I. Coupling near the plasma level}",
      journal = {Australian Journal of Physics},
         year = 1974,
        month = feb,
       volume = {27},
        pages = {31},
          doi = {10.1071/PH740031},
       adsurl = {https://ui.adsabs.harvard.edu/abs/1974AuJPh..27...31M}
}

@ARTICLE{Melrose_2006,
       author = {{Melrose}, D.~B.},
        title = "{Depolarization of Radio Bursts Due to Reflection off Sharp Boundaries in the Solar Corona}",
      journal = {\apj},
         year = 2006,
        month = feb,
       volume = {637},
       number = {2},
        pages = {1113-1121},
          doi = {10.1086/498499},
archivePrefix = {arXiv},
       eprint = {astro-ph/0507531},
 primaryClass = {astro-ph},
       adsurl = {https://ui.adsabs.harvard.edu/abs/2006ApJ...637.1113M}
}

@ARTICLE{Maguire2020,
       author = {{Maguire}, Ciara A. and {Carley}, Eoin P. and {McCauley}, Joseph and {Gallagher}, Peter T.},
        title = "{Evolution of the Alfv{\'e}n Mach number associated with a coronal mass ejection shock}",
      journal = {\aap},
         year = 2020,
        month = jan,
       volume = {633},
          eid = {A56},
        pages = {A56},
          doi = {10.1051/0004-6361/201936449},
archivePrefix = {arXiv},
       eprint = {1912.01863},
 primaryClass = {astro-ph.SR},
       adsurl = {https://ui.adsabs.harvard.edu/abs/2020A&A...633A..56M}
}

@INCOLLECTION{Nelson_Melrose1985_typeII_burst,
       author = {{Nelson}, G.~J. and {Melrose}, D.~B.},
        title = "{Type II bursts.}",
    booktitle = {Solar Radiophysics: Studies of Emission from the Sun at Metre Wavelengths},
         year = 1985,
       editor = {{McLean}, D.~J. and {Labrum}, N.~R.},
        pages = {333-359},
       adsurl = {https://ui.adsabs.harvard.edu/abs/1985srph.book..333N}
}

@ARTICLE{Melrose1980,
       author = {{Melrose}, D.~B. and {Dulk}, G.~A. and {Gary}, D.~E.},
        title = "{Corrected formula for the polarization of second harmonic plasma emission}",
      journal = {\pasa},
         year = 1980,
        month = jan,
       volume = {4},
       number = {1},
        pages = {50-53},
          doi = {10.1017/S1323358000018762},
       adsurl = {https://ui.adsabs.harvard.edu/abs/1980PASA....4...50M}
}

@ARTICLE{Zlotnik1981,
       author = {{Zlotnik}, E. Ia.},
        title = "{The polarization of second harmonic radio emission in Type III bursts}",
      journal = {\aap},
         year = 1981,
        month = aug,
       volume = {101},
       number = {2},
        pages = {250-258},
       adsurl = {https://ui.adsabs.harvard.edu/abs/1981A&A...101..250Z}
}

@ARTICLE{Komesaroff1958,
       author = {{Komesaroff}, M.},
        title = "{Polarization Measurements of the Three Spectral Types of Solar Radio Burst}",
      journal = {Australian Journal of Physics},
         year = 1958,
        month = jun,
       volume = {11},
        pages = {201},
          doi = {10.1071/PH580201},
       adsurl = {https://ui.adsabs.harvard.edu/abs/1958AuJPh..11..201K}
}

@ARTICLE{Melrose1977,
       author = {{Melrose}, D.~V.},
        title = "{Remarks on plasma emission and its application to solar radio bursts.}",
      journal = {Izvestiia Vysshaia Uchebn. Zaved., Radiofizika},
         year = 1977,
        month = jan,
       volume = {20},
        pages = {1369-1378},
       adsurl = {https://ui.adsabs.harvard.edu/abs/1977IzVUZ..20.1369M}
}

@ARTICLE{MelroseSy1972,
       author = {{Melrose}, D.~B. and {Sy}, W.~N.},
        title = "{Plasma emission processes in a magnetoactive plasma}",
      journal = {Australian Journal of Physics},
         year = 1972,
        month = aug,
       volume = {25},
        pages = {387},
          doi = {10.1071/PH720387},
       adsurl = {https://ui.adsabs.harvard.edu/abs/1972AuJPh..25..387M}
}

@ARTICLE{Alissandrakis2021,
       author = {{Alissandrakis}, C.~E. and {Nindos}, A. and {Patsourakos}, S. and {Hillaris}, A.},
        title = "{Multiwavelength observations of a metric type-II event}",
      journal = {\aap},
         year = 2021,
        month = oct,
       volume = {654},
          eid = {A112},
        pages = {A112},
          doi = {10.1051/0004-6361/202141672},
archivePrefix = {arXiv},
       eprint = {2108.02855},
 primaryClass = {astro-ph.SR},
       adsurl = {https://ui.adsabs.harvard.edu/abs/2021A&A...654A.112A}
}

@ARTICLE{Thejappa2003,
       author = {{Thejappa}, G. and {Zlobec}, P. and {MacDowall}, R.~J.},
        title = "{Polarization and Fragmentation of Solar Type II Radio Bursts}",
      journal = {\apj},
         year = 2003,
        month = aug,
       volume = {592},
       number = {2},
        pages = {1234-1240},
          doi = {10.1086/375859},
       adsurl = {https://ui.adsabs.harvard.edu/abs/2003ApJ...592.1234T}
}

@ARTICLE{Morosan2022,
       author = {{Morosan}, Diana E. and {R{\"a}s{\"a}nen}, Juska E. and {Kumari}, Anshu and {Kilpua}, Emilia K.~J. and {Bisi}, Mario M. and {Dabrowski}, Bartosz and {Krankowski}, Andrzej and {Magdaleni{\'c}}, Jasmina and {Mann}, Gottfried and {Rothkaehl}, Hanna and {Vocks}, Christian and {Zucca}, Pietro},
        title = "{Exploring the Circular Polarisation of Low-Frequency Solar Radio Bursts with LOFAR}",
      journal = {\solphys},
         year = 2022,
        month = apr,
       volume = {297},
       number = {4},
          eid = {47},
        pages = {47},
          doi = {10.1007/s11207-022-01976-9},
archivePrefix = {arXiv},
       eprint = {2203.14674},
 primaryClass = {astro-ph.SR},
       adsurl = {https://ui.adsabs.harvard.edu/abs/2022SoPh..297...47M}
}

@article{Kumari2017typeII_band,
 adsurl = {https://ui.adsabs.harvard.edu/abs/2017SoPh..292..161K},
 archiveprefix = {arXiv},
 author = {{Kumari}, Anshu and {Ramesh}, R. and {Kathiravan}, C. and {Wang}, T.~J.},
 doi = {10.1007/s11207-017-1180-6},
 eid = {161},
 eprint = {1711.02307},
 journal = {\solphys},
 month = {November},
 number = {11},
 pages = {161},
 primaryclass = {astro-ph.SR},
 title = {{Strength of the Solar Coronal Magnetic Field - A Comparison of Independent Estimates Using Contemporaneous Radio and White-Light Observations}},
 volume = {292},
 year = {2017}
}

@ARTICLE{Mondal2024,
       author = {{Mondal}, Surajit and {Kansabanik}, Devojyoti and {Oberoi}, Divya and {Dey}, Soham},
        title = "{New Insights into Type-I Solar Noise Storms from High Angular Resolution Spectroscopic Imaging with the Upgraded Giant Metrewave Radio Telescope}",
      journal = {\apj},
         year = 2024,
        month = nov,
       volume = {975},
       number = {1},
          eid = {122},
        pages = {122},
          doi = {10.3847/1538-4357/ad77c2},
archivePrefix = {arXiv},
       eprint = {2408.13465},
 primaryClass = {astro-ph.SR},
       adsurl = {https://ui.adsabs.harvard.edu/abs/2024ApJ...975..122M}
}

@ARTICLE{Mondal2025,
       author = {{Mondal}, Surajit and {Zhang}, Peijin and {Kansabanik}, Devojyoti and {Oberoi}, Divya and {Pearce}, Gillian},
        title = "{Observation and Modeling of Small Spatial Structures of Solar Radio Noise Storms Using the uGMRT}",
      journal = {\solphys},
         year = 2025,
        month = jul,
       volume = {300},
       number = {8},
          eid = {109},
        pages = {109},
          doi = {10.1007/s11207-025-02519-8},
archivePrefix = {arXiv},
       eprint = {2507.08093},
 primaryClass = {astro-ph.SR},
       adsurl = {https://ui.adsabs.harvard.edu/abs/2025SoPh..300..109M}
}

@ARTICLE{Callingham2021,
       author = {{Callingham}, J.~R. and {Pope}, B.~J.~S. and {Feinstein}, A.~D. and {Vedantham}, H.~K. and {Shimwell}, T.~W. and {Zarka}, P. and {Tasse}, C. and {Lamy}, L. and {Veken}, K. and {Toet}, S. and {Sabater}, J. and {Best}, P.~N. and {van Weeren}, R.~J. and {R{\"o}ttgering}, H.~J.~A. and {Ray}, T.~P.},
        title = "{Low-frequency monitoring of flare star binary CR Draconis: long-term electron-cyclotron maser emission}",
      journal = {\aap},
         year = 2021,
        month = apr,
       volume = {648},
          eid = {A13},
        pages = {A13},
          doi = {10.1051/0004-6361/202039144},
archivePrefix = {arXiv},
       eprint = {2102.04751},
 primaryClass = {astro-ph.SR},
       adsurl = {https://ui.adsabs.harvard.edu/abs/2021A&A...648A..13C}
}

@ARTICLE{Bastian2022,
       author = {{Bastian}, T.~S. and {Cotton}, W.~D. and {Hallinan}, G.},
        title = "{Radio Emission from UV Cet: Auroral Emission from a Stellar Magnetosphere}",
      journal = {\apj},
         year = 2022,
        month = aug,
       volume = {935},
       number = {2},
          eid = {99},
        pages = {99},
          doi = {10.3847/1538-4357/ac7d57},
archivePrefix = {arXiv},
       eprint = {2206.14099},
 primaryClass = {astro-ph.SR},
       adsurl = {https://ui.adsabs.harvard.edu/abs/2022ApJ...935...99B}
}

@ARTICLE{Dulk_McLean_1978,
       author = {{Dulk}, G.~A. and {McLean}, D.~J.},
        title = "{Coronal magnetic fields.}",
      journal = {\solphys},
         year = 1978,
        month = apr,
       volume = {57},
       number = {2},
        pages = {279-295},
          doi = {10.1007/BF00160102},
       adsurl = {https://ui.adsabs.harvard.edu/abs/1978SoPh...57..279D}
}

@ARTICLE{Callingham2025_stellar_typeII,
       author = {{Callingham}, J.~R. and {Tasse}, C. and {Keers}, R. and {Kavanagh}, R.~D. and {Vedantham}, H.~K. and {Zarka}, P. and {Bellotti}, S. and {Cristofari}, P.~I. and {Bloot}, S. and {Konijn}, D.~C. and {Hardcastle}, M.~J. and {Lamy}, L. and {Pass}, E.~K. and {Pope}, B.~J.~S. and {Reid}, H. and {R{\"o}ttgering}, H.~J.~A. and {Shimwell}, T.~W. and {Zucca}, P.},
        title = "{Radio burst from a stellar coronal mass ejection}",
      journal = {\nat},
         year = 2025,
        month = nov,
       volume = {647},
       number = {8090},
        pages = {603-607},
          doi = {10.1038/s41586-025-09715-3},
archivePrefix = {arXiv},
       eprint = {2511.09289},
 primaryClass = {astro-ph.SR},
       adsurl = {https://ui.adsabs.harvard.edu/abs/2025Natur.647..603C}
}

@ARTICLE{Dey_2025_linear,
       author = {{Dey}, Soham and {Kansabanik}, Devojyoti and {Oberoi}, Divya and {Mondal}, Surajit},
        title = "{First Robust Detection of Linear Polarization from Metric Solar Emissions: Challenging Established Paradigms}",
      journal = {\apjl},
         year = 2025,
        month = aug,
       volume = {988},
       number = {2},
          eid = {L73},
        pages = {L73},
          doi = {10.3847/2041-8213/adef0e},
archivePrefix = {arXiv},
       eprint = {2507.05196},
 primaryClass = {astro-ph.SR},
       adsurl = {https://ui.adsabs.harvard.edu/abs/2025ApJ...988L..73D}
}

@ARTICLE{sunpy_community2020,
  doi = {10.3847/1538-4357/ab4f7a},
  url = {https://iopscience.iop.org/article/10.3847/1538-4357/ab4f7a},
  author = {{The SunPy Community} and Barnes, Will T. and Bobra, Monica G. and Christe, Steven D. and Freij, Nabil and Hayes, Laura A. and Ireland, Jack and Mumford, Stuart and Perez-Suarez, David and Ryan, Daniel F. and Shih, Albert Y. and Chanda, Prateek and Glogowski, Kolja and Hewett, Russell and Hughitt, V. Keith and Hill, Andrew and Hiware, Kaustubh and Inglis, Andrew and Kirk, Michael S. F. and Konge, Sudarshan and Mason, James Paul and Maloney, Shane Anthony and Murray, Sophie A. and Panda, Asish and Park, Jongyeob and Pereira, Tiago M. D. and Reardon, Kevin and Savage, Sabrina and Sipőcz, Brigitta M. and Stansby, David and Jain, Yash and Taylor, Garrison and Yadav, Tannmay and Rajul and Dang, Trung Kien},
  title = {The SunPy Project: Open Source Development and Status of the Version 1.0 Core Package},
  journal = {The Astrophysical Journal},
  volume = {890},
  issue = {1},
  pages = {68-},
  publisher = {American Astronomical Society},
  year = {2020}
}

@ARTICLE{Bastian1995,
       author = {{Bastian}, T.~S.},
        title = "{Angular Scattering of Radio Waves: Implications for Mode Coupling in the Solar Corona}",
      journal = {\apj},
         year = 1995,
        month = jan,
       volume = {439},
        pages = {494},
          doi = {10.1086/175190},
       adsurl = {https://ui.adsabs.harvard.edu/abs/1995ApJ...439..494B}
}

@ARTICLE{cohen1960,
       author = {{Cohen}, M.~H.},
        title = "{Magnetoionic Mode Coupling at High Frequencies.}",
      journal = {\apj},
         year = 1960,
        month = may,
       volume = {131},
        pages = {664},
          doi = {10.1086/146878},
       adsurl = {https://ui.adsabs.harvard.edu/abs/1960ApJ...131..664C}
}

@ARTICLE{Piddington1951,
       author = {{Piddington}, J.~H. and {Minnett}, H.~C.},
        title = "{Solar Radio-Frequency Emission from Localized Regions at Very High Temperatures}",
      journal = {Australian Journal of Scientific Research A Physical Sciences},
         year = 1951,
        month = jun,
       volume = {4},
        pages = {131},
          doi = {10.1071/CH9510131},
       adsurl = {https://ui.adsabs.harvard.edu/abs/1951AuSRA...4..131P}
}

@ARTICLE{Gopalswamy1994,
       author = {{Gopalswamy}, N. and {Zheleznyakov}, V.~V. and {White}, S.~M. and {Kundu}, M.~R.},
        title = "{Polarization Features of Solar Radio Emission and Possible Existence of Current Sheets in Active Regions}",
      journal = {\solphys},
         year = 1994,
        month = dec,
       volume = {155},
       number = {2},
        pages = {339-350},
          doi = {10.1007/BF00680599},
       adsurl = {https://ui.adsabs.harvard.edu/abs/1994SoPh..155..339G}
}

@Article{         harris2020array,
 title         = {Array programming with {NumPy}},
 author        = {Charles R. Harris and K. Jarrod Millman and St{\'{e}}fan J.
                 van der Walt and Ralf Gommers and Pauli Virtanen and David
                 Cournapeau and Eric Wieser and Julian Taylor and Sebastian
                 Berg and Nathaniel J. Smith and Robert Kern and Matti Picus
                 and Stephan Hoyer and Marten H. van Kerkwijk and Matthew
                 Brett and Allan Haldane and Jaime Fern{\'{a}}ndez del
                 R{\'{i}}o and Mark Wiebe and Pearu Peterson and Pierre
                 G{\'{e}}rard-Marchant and Kevin Sheppard and Tyler Reddy and
                 Warren Weckesser and Hameer Abbasi and Christoph Gohlke and
                 Travis E. Oliphant},
 year          = {2020},
 month         = sep,
 journal       = {Nature},
 volume        = {585},
 number        = {7825},
 pages         = {357--362},
 doi           = {10.1038/s41586-020-2649-2},
 publisher     = {Springer Science and Business Media {LLC}},
 url           = {https://doi.org/10.1038/s41586-020-2649-2}
}

@ARTICLE{astropy2022,
       author = {{Astropy Collaboration} and {Price-Whelan}, Adrian M. and {Lim}, Pey Lian and {Earl}, Nicholas and {Starkman}, Nathaniel and {Bradley}, Larry and {Shupe}, David L. and {Patil}, Aarya A. and {Corrales}, Lia and {Brasseur}, C.~E. and {N{\"o}the}, Maximilian and {Donath}, Axel and {Tollerud}, Erik and {Morris}, Brett M. and {Ginsburg}, Adam and {Vaher}, Eero and {Weaver}, Benjamin A. and {Tocknell}, James and {Jamieson}, William and {van Kerkwijk}, Marten H. and {Robitaille}, Thomas P. and {Merry}, Bruce and {Bachetti}, Matteo and {G{\"u}nther}, H. Moritz and {Aldcroft}, Thomas L. and {Alvarado-Montes}, Jaime A. and {Archibald}, Anne M. and {B{\'o}di}, Attila and {Bapat}, Shreyas and {Barentsen}, Geert and {Baz{\'a}n}, Juanjo and {Biswas}, Manish and {Boquien}, M{\'e}d{\'e}ric and {Burke}, D.~J. and {Cara}, Daria and {Cara}, Mihai and {Conroy}, Kyle E. and {Conseil}, Simon and {Craig}, Matthew W. and {Cross}, Robert M. and {Cruz}, Kelle L. and {D'Eugenio}, Francesco and {Dencheva}, Nadia and {Devillepoix}, Hadrien A.~R. and {Dietrich}, J{\"o}rg P. and {Eigenbrot}, Arthur Davis and {Erben}, Thomas and {Ferreira}, Leonardo and {Foreman-Mackey}, Daniel and {Fox}, Ryan and {Freij}, Nabil and {Garg}, Suyog and {Geda}, Robel and {Glattly}, Lauren and {Gondhalekar}, Yash and {Gordon}, Karl D. and {Grant}, David and {Greenfield}, Perry and {Groener}, Austen M. and {Guest}, Steve and {Gurovich}, Sebastian and {Handberg}, Rasmus and {Hart}, Akeem and {Hatfield-Dodds}, Zac and {Homeier}, Derek and {Hosseinzadeh}, Griffin and {Jenness}, Tim and {Jones}, Craig K. and {Joseph}, Prajwel and {Kalmbach}, J. Bryce and {Karamehmetoglu}, Emir and {Ka{\l}uszy{\'n}ski}, Miko{\l}aj and {Kelley}, Michael S.~P. and {Kern}, Nicholas and {Kerzendorf}, Wolfgang E. and {Koch}, Eric W. and {Kulumani}, Shankar and {Lee}, Antony and {Ly}, Chun and {Ma}, Zhiyuan and {MacBride}, Conor and {Maljaars}, Jakob M. and {Muna}, Demitri and {Murphy}, N.~A. and {Norman}, Henrik and {O'Steen}, Richard and {Oman}, Kyle A. and {Pacifici}, Camilla and {Pascual}, Sergio and {Pascual-Granado}, J. and {Patil}, Rohit R. and {Perren}, Gabriel I. and {Pickering}, Timothy E. and {Rastogi}, Tanuj and {Roulston}, Benjamin R. and {Ryan}, Daniel F. and {Rykoff}, Eli S. and {Sabater}, Jose and {Sakurikar}, Parikshit and {Salgado}, Jes{\'u}s and {Sanghi}, Aniket and {Saunders}, Nicholas and {Savchenko}, Volodymyr and {Schwardt}, Ludwig and {Seifert-Eckert}, Michael and {Shih}, Albert Y. and {Jain}, Anany Shrey and {Shukla}, Gyanendra and {Sick}, Jonathan and {Simpson}, Chris and {Singanamalla}, Sudheesh and {Singer}, Leo P. and {Singhal}, Jaladh and {Sinha}, Manodeep and {Sip{\H{o}}cz}, Brigitta M. and {Spitler}, Lee R. and {Stansby}, David and {Streicher}, Ole and {{\v{S}}umak}, Jani and {Swinbank}, John D. and {Taranu}, Dan S. and {Tewary}, Nikita and {Tremblay}, Grant R. and {de Val-Borro}, Miguel and {Van Kooten}, Samuel J. and {Vasovi{\'c}}, Zlatan and {Verma}, Shresth and {de Miranda Cardoso}, Jos{\'e} Vin{\'\i}cius and {Williams}, Peter K.~G. and {Wilson}, Tom J. and {Winkel}, Benjamin and {Wood-Vasey}, W.~M. and {Xue}, Rui and {Yoachim}, Peter and {Zhang}, Chen and {Zonca}, Andrea and {Astropy Project Contributors}},
        title = "{The Astropy Project: Sustaining and Growing a Community-oriented Open-source Project and the Latest Major Release (v5.0) of the Core Package}",
      journal = {\apj},
         year = 2022,
        month = aug,
       volume = {935},
       number = {2},
          eid = {167},
        pages = {167},
          doi = {10.3847/1538-4357/ac7c74},
archivePrefix = {arXiv},
       eprint = {2206.14220},
 primaryClass = {astro-ph.IM},
       adsurl = {https://ui.adsabs.harvard.edu/abs/2022ApJ...935..167A}
}

@ARTICLE{astropy2018,
       author = {{Astropy Collaboration} and {Price-Whelan}, A.~M. and {Sip{\H{o}}cz}, B.~M. and {G{\"u}nther}, H.~M. and {Lim}, P.~L. and {Crawford}, S.~M. and {Conseil}, S. and {Shupe}, D.~L. and {Craig}, M.~W. and {Dencheva}, N. and {Ginsburg}, A. and {VanderPlas}, J.~T. and {Bradley}, L.~D. and {P{\'e}rez-Su{\'a}rez}, D. and {de Val-Borro}, M. and {Aldcroft}, T.~L. and {Cruz}, K.~L. and {Robitaille}, T.~P. and {Tollerud}, E.~J. and {Ardelean}, C. and {Babej}, T. and {Bach}, Y.~P. and {Bachetti}, M. and {Bakanov}, A.~V. and {Bamford}, S.~P. and {Barentsen}, G. and {Barmby}, P. and {Baumbach}, A. and {Berry}, K.~L. and {Biscani}, F. and {Boquien}, M. and {Bostroem}, K.~A. and {Bouma}, L.~G. and {Brammer}, G.~B. and {Bray}, E.~M. and {Breytenbach}, H. and {Buddelmeijer}, H. and {Burke}, D.~J. and {Calderone}, G. and {Cano Rodr{\'\i}guez}, J.~L. and {Cara}, M. and {Cardoso}, J.~V.~M. and {Cheedella}, S. and {Copin}, Y. and {Corrales}, L. and {Crichton}, D. and {D'Avella}, D. and {Deil}, C. and {Depagne}, {\'E}. and {Dietrich}, J.~P. and {Donath}, A. and {Droettboom}, M. and {Earl}, N. and {Erben}, T. and {Fabbro}, S. and {Ferreira}, L.~A. and {Finethy}, T. and {Fox}, R.~T. and {Garrison}, L.~H. and {Gibbons}, S.~L.~J. and {Goldstein}, D.~A. and {Gommers}, R. and {Greco}, J.~P. and {Greenfield}, P. and {Groener}, A.~M. and {Grollier}, F. and {Hagen}, A. and {Hirst}, P. and {Homeier}, D. and {Horton}, A.~J. and {Hosseinzadeh}, G. and {Hu}, L. and {Hunkeler}, J.~S. and {Ivezi{\'c}}, {\v{Z}}. and {Jain}, A. and {Jenness}, T. and {Kanarek}, G. and {Kendrew}, S. and {Kern}, N.~S. and {Kerzendorf}, W.~E. and {Khvalko}, A. and {King}, J. and {Kirkby}, D. and {Kulkarni}, A.~M. and {Kumar}, A. and {Lee}, A. and {Lenz}, D. and {Littlefair}, S.~P. and {Ma}, Z. and {Macleod}, D.~M. and {Mastropietro}, M. and {McCully}, C. and {Montagnac}, S. and {Morris}, B.~M. and {Mueller}, M. and {Mumford}, S.~J. and {Muna}, D. and {Murphy}, N.~A. and {Nelson}, S. and {Nguyen}, G.~H. and {Ninan}, J.~P. and {N{\"o}the}, M. and {Ogaz}, S. and {Oh}, S. and {Parejko}, J.~K. and {Parley}, N. and {Pascual}, S. and {Patil}, R. and {Patil}, A.~A. and {Plunkett}, A.~L. and {Prochaska}, J.~X. and {Rastogi}, T. and {Reddy Janga}, V. and {Sabater}, J. and {Sakurikar}, P. and {Seifert}, M. and {Sherbert}, L.~E. and {Sherwood-Taylor}, H. and {Shih}, A.~Y. and {Sick}, J. and {Silbiger}, M.~T. and {Singanamalla}, S. and {Singer}, L.~P. and {Sladen}, P.~H. and {Sooley}, K.~A. and {Sornarajah}, S. and {Streicher}, O. and {Teuben}, P. and {Thomas}, S.~W. and {Tremblay}, G.~R. and {Turner}, J.~E.~H. and {Terr{\'o}n}, V. and {van Kerkwijk}, M.~H. and {de la Vega}, A. and {Watkins}, L.~L. and {Weaver}, B.~A. and {Whitmore}, J.~B. and {Woillez}, J. and {Zabalza}, V. and {Astropy Contributors}},
        title = "{The Astropy Project: Building an Open-science Project and Status of the v2.0 Core Package}",
      journal = {\aj},
         year = 2018,
        month = sep,
       volume = {156},
       number = {3},
          eid = {123},
        pages = {123},
          doi = {10.3847/1538-3881/aabc4f},
archivePrefix = {arXiv},
       eprint = {1801.02634},
 primaryClass = {astro-ph.IM},
       adsurl = {https://ui.adsabs.harvard.edu/abs/2018AJ....156..123A}
}

@Article{Hunter:2007,
  Author    = {Hunter, J. D.},
  Title     = {Matplotlib: A 2D graphics environment},
  Journal   = {Computing in Science \& Engineering},
  Volume    = {9},
  Number    = {3},
  Pages     = {90--95},
  doi       = {10.1109/MCSE.2007.55},
  year      = 2007
}

@ARTICLE{astropy2013,
       author = {{Astropy Collaboration} and {Robitaille}, Thomas P. and {Tollerud}, Erik J. and {Greenfield}, Perry and {Droettboom}, Michael and {Bray}, Erik and {Aldcroft}, Tom and {Davis}, Matt and {Ginsburg}, Adam and {Price-Whelan}, Adrian M. and {Kerzendorf}, Wolfgang E. and {Conley}, Alexander and {Crighton}, Neil and {Barbary}, Kyle and {Muna}, Demitri and {Ferguson}, Henry and {Grollier}, Fr{\'e}d{\'e}ric and {Parikh}, Madhura M. and {Nair}, Prasanth H. and {Unther}, Hans M. and {Deil}, Christoph and {Woillez}, Julien and {Conseil}, Simon and {Kramer}, Roban and {Turner}, James E.~H. and {Singer}, Leo and {Fox}, Ryan and {Weaver}, Benjamin A. and {Zabalza}, Victor and {Edwards}, Zachary I. and {Azalee Bostroem}, K. and {Burke}, D.~J. and {Casey}, Andrew R. and {Crawford}, Steven M. and {Dencheva}, Nadia and {Ely}, Justin and {Jenness}, Tim and {Labrie}, Kathleen and {Lim}, Pey Lian and {Pierfederici}, Francesco and {Pontzen}, Andrew and {Ptak}, Andy and {Refsdal}, Brian and {Servillat}, Mathieu and {Streicher}, Ole},
        title = "{Astropy: A community Python package for astronomy}",
      journal = {\aap},
         year = 2013,
        month = oct,
       volume = {558},
          eid = {A33},
        pages = {A33},
          doi = {10.1051/0004-6361/201322068},
archivePrefix = {arXiv},
       eprint = {1307.6212},
 primaryClass = {astro-ph.IM},
       adsurl = {https://ui.adsabs.harvard.edu/abs/2013A&A...558A..33A}
}

@article{sullivan2019pyvista,
  doi = {10.21105/joss.01450},
  url = {https://doi.org/10.21105/joss.01450},
  year = {2019},
  month = {May},
  publisher = {The Open Journal},
  volume = {4},
  number = {37},
  pages = {1450},
  author = {Bane Sullivan and Alexander Kaszynski},
  title = {{PyVista}: {3D} plotting and mesh analysis through a streamlined interface for the {Visualization Toolkit} ({VTK})},
  journal = {Journal of Open Source Software}
}
\bibliographystyle{aasjournal}

\end{document}